\documentclass[manuscript]{aastex}
\shortauthors{Kim et al.}
\slugcomment{\today}

\begin{document}
\title{Near-Infrared Polarimetric Study of the N159/N160 Star-Forming Complex in the Large Magellanic Cloud}

\author{Jaeyeong Kim\altaffilmark{1}$^{,}$\altaffilmark{2}
        Woong-Seob Jeong\altaffilmark{2}$^{,}$\altaffilmark{3}$^{,}$\altaffilmark{8}\footnotetext[8]{Corresponding author},
        Jeonghyun Pyo\altaffilmark{2}, Soojong Pak\altaffilmark{1}, Won-Kee Park\altaffilmark{2},
        Jungmi Kwon\altaffilmark{4,5,6}, and Motohide Tamura\altaffilmark{7}}
\affil{$^1$ School of Space Research, Kyung Hee University,
            1732 Deogyeong-daero, Giheung-gu, Yongin, Gyeonggi-do 17104,
            Republic of Korea; jaeyeong@khu.ac.kr}
\affil{$^2$ Korea Astronomy and Space Science Institute,
            776 Daedeok-daero, Yuseong-gu, Daejeon 34055,
            Republic of Korea; jeongws@kasi.re.kr}
\affil{$^3$ Korea University of Science and Technology,
            217 Gajeong-ro, Yuseong-gu, Daejeon 34113,
            Republic of Korea}
\affil{$^4$ Department of Astronomy, Graduate School of Science, The University of Tokyo,
            7-3-1 Hongo, Bunkyo-ku, Tokyo 113-0033,
            Japan}
\affil{$^5$ National Astronomical Observatory of Japan,
            2-21-1 Osawa, Mitaka, Tokyo 181-8588,
            Japan}
\affil{$^6$ Institute of Space and Astronautical Science, Japan Aerospace Exploration Agency,
            3-1-1 Yoshinodai, Chuo-ku, Sagamihara, Kanagawa 252-5210,
            Japan}
\affil{$^7$ The University of Tokyo/National Astronomical Observatory of Japan/Astrobiology Center,
            2-21-1 Osawa, Mitaka, Tokyo 181-8588,
            Japan}

\begin{abstract}
We present near-infrared polarimetric results for the N159/N160 star-forming complex in the Large Magellanic Cloud (LMC) with SIRPOL, the polarimeter of the Infrared Survey Facility.
We separated foreground sources using their visual extinction derived from near-infrared photometric data.
The 45 young stellar candidates and 2 high-excitation blobs were matched with our sources, and 12 of them showed the high polarization that did not originate from the interstellar dust.
We made a polarimetric catalog of 252, 277, and 89 sources at the $J$, $H$, and $K_s$ bands, respectively.
Based on the ratios of the polarization degree between these bands, we verify that the origin of these polarized sources is the dichroic extinction from the interstellar dust aligned by the magnetic field and that the ratios follow a power-law dependence of $P_{\lambda}$ $\sim$ $\lambda^{-0.9}$.
The linear polarization vectors projected onto the H$\alpha$ image of the complex turned out to follow the local magnetic field structure.
The vector map overlaid on dust and gas emissions shows the close correlation between the magnetic field structure and surrounding interstellar medium.
We suggest that the derived magnetic field structure supports the sequential formation scenario of the complex.
\end{abstract}


\section{Introduction}
It is important to understand the formation and evolution of stars in molecular clouds, because they are the most fundamental building blocks of the universe.
Magnetic fields have been considered as one of the main forces influencing the star formation process, thus there have been many studies on them both theoretically and observationally \citep{shu87,hei05,mc07,cru12,dob14}.
\citet{li14} reviewed the observations of magnetic fields by their scales, and discussed the relation between magnetic fields and the formation process of molecular clouds and stars.
Since magnetic fields are generally associated with the interstellar medium, revealing their structure of magnetic fields is essential to understand the morphology of star-forming regions and their surrounding interstellar medium.

Polarimetry is the general method for examining magnetic fields.
Near-infrared polarimetric measurements in star-forming regions provide us with the magnetic field structure by observing the dichroic extinction of background stars, whose starlights are polarized when they pass through the magnetically aligned dust grains within molecular clouds.
Various polarimetric studies of magnetic fields have been performed in Galactic star-forming regions \citep{kan07,tam07,ku08,sai09,kwon11,san14,kus15}.

The Large Magellanic Cloud (LMC) is one of the nearest galaxies, and therefore it has been one of the best observed targets in studies of the star-forming regions and giant molecular clouds (GMCs) within a galaxy.
The northeastern part of the LMC includes the root of the molecular stream extending southward from the 30 Doradus complex and has been the focus of numerous studies of the star-forming regions (i.e. the 30 Doradus, N158, and N159/N160 complexes) and molecular clouds \citep{coh88,kut97,fuk99,ott08,pin09}.
\citet{nak07} first studied the central part of 30 Doradus using near-infrared polarimetric observations, and discovered its magnetic field distribution, which showed a U-shaped structure in the northern area and aligned about 70$\arcdeg$ and 45$\arcdeg$ in the western and southeastern areas, respectively.
\citet{kim11} observed a wider region of 20$\arcmin$ $\times$ 20$\arcmin$ region around 30 Doradus, and they used the proper motion data to separate the Galactic foreground sources.
They confirmed that most polarized sources showed their polarization angles of 75$\arcdeg$.
\citet{kim16} extended the survey area to 39$\arcmin$ $\times$ 69$\arcmin$, and found patterns following the molecular stream.
However, the N159/N160 complex still shows the veiled structure of the magnetic fields.
Since this region has active star formation, \ion{H}{2} regions, and interesting molecular cloud structures (see Section \ref{sec:N159/N160} for details), we expect that the magnetic fields might be influenced by those environments to make construct complex structure.

This research studied the detailed structure of the magnetic field in the N159/N160 complex.
We carried out deep near-infrared polarimetric observations over the two fields of the complex.
In Section 2, we describe the N159/N160 complex and review previous studies of this region.
Observations and data reduction are described in Section 3.
The method to separate foreground sources is also presented in Section 3.
The polarimetric results and analysis are presented in Section 4.
In Section 5, we discuss the magnetic field structures in the N159/N160 complex with gas and dust distributions, and suggest a star formation scenario in the complex.

\section{The N159/N160 Complex}\label{sec:N159/N160}
The N159/N160 complex is located in a molecular ridge, about 600 pc south from the 30 Doradus complex.
Ever since \citet{hen56} cataloged \ion{H}{2} regions from the survey of emission nebulae in the LMC, this complex has been the target of various observations \citep{jon86,joh98,pak98,bol00,jon05,nak05,miz10,car12,gal13}.
The N159/N160 complex is composed of the N160 and N159 \ion{H}{2} regions, and they are spatially separated from each other in the northern and the southern parts of the complex.
The H$\alpha$ observations by the Wide Field Imager (WFI) of the ESO 2.2 m telescope (program ID 076.C-0888\footnote{Data are available and described at http://archive.eso.org/cms/eso-data/data-packages/30-doradus/30-doradus-reqpage.html.}) clearly show the detailed structure of the two regions.

N160 includes \ion{H}{2} regions and young stellar clusters.
Bright \ion{H}{2} regions in N160 are designated using the notation of \citet{hen56}.
The main \ion{H}{2} region, with an elongated shape, is composed of N160 A and N160 D (hereafter N160 A+D) at the southwestern part of N160.
The southern part of N160 has two bright condensations of \ion{H}{2} regions, N160 B and N160 C (hereafter N160 B+C).
\citet{hey86} observed the \ion{H}{2} regions in N160 and reported the physical properties of high-excitation blobs (HEBs) and their surrounding nebulae.
HEBs are a rare class of the compact \ion{H}{2} regions in the Magellanic Clouds.
\citet{mar08} observed HEBs in N160 A with high spatial resolution.
A notable feature of the N160 region is a shell structure opened toward the northeast.
The parent clouds of the shell are almost dissipated and young clusters are detected inside the shell \citep{nak05}.
Young stellar objects (YSOs) have been observed around the \ion{H}{2} regions in N160 \citep{henn98,nak05,van10,car12}.
Sequential star formation throughout this region was suggested, based on the various evolutionary stages of massive stars in the OB clusters \citep{nak05,far09}.

N159 has bright \ion{H}{2} regions and two GMCs, N159E and N159W, located at the eastern and southwestern parts of the region, respectively.
In addition, N159 has embedded interesting objects such as an X-ray binary, a supernova remnant, compact \ion{H}{2} regions, and radio sources \citep{hunt94,cow95,chu97,hey99,tes06,tes07,seok13}.
\citet{jon05} located these objects in mid-infrared images observed with the $Spitzer$ $Space$ $Telescope$.
Studies of dust emissions unveiled that N159 is younger than N160 and shows active star formation \citep{gal13}.
CO observations detected the sites of star formation through cloud-cloud collisions, and revealed the link between high concentrations of molecular gas and star formation \citep{miz10,fuk15}.
Embedded massive YSOs, maser sources, and ultracompact \ion{H}{2} regions are distributed around the molecular clouds \citep{nak05,che10}.

\section{Observations and Data Reduction}

\subsection{Observations}
Simultaneous $JHK_s$ polarimetric observations of the N159/N160 complex were performed on 2007 February 3 and 5.
We used the near-infrared camera SIRIUS \citep{nag03} and the polarimeter SIRPOL \citep{kan06} of the Infrared Survey Facility (IRSF) 1.4 m telescope at the South African Astronomical Observatory in Sutherland, South Africa.
The camera has a field of view of 7\farcm7 $\times$ 7\farcm7 and a pixel scale of 0\farcs45 pixel$^{-1}$.
One set of observations for a target field consisted of 20 s exposures at 10 dithered positions for four wave-plate angles (0$\arcdeg$, 45$\arcdeg$, 22\fdg5, and 67\fdg5) in the $J$, $H$, and $K_s$ bands, and the whole sequence is repeated 10 and 9 times for the N159 and N160 fields centered at ($\alpha$, $\delta$)$_{2000}$ $=$ (5\textsuperscript{h}39\textsuperscript{m}37\textsuperscript{s}.1, -69$\arcdeg$43$\arcmin$45\farcs1) and (5\textsuperscript{h}40\textsuperscript{m}05\textsuperscript{s}.6, -69$\arcdeg$36$\arcmin$25\farcs8), respectively.

\subsection{Data Reduction}
We used the DAOPHOT package \citep{ste87} of the Image Reduction \& Analysis Facility\footnote{IRAF is distributed by the US National Optical Astronomy Observatories, which are operated by the Association of Universities for Research in Astronomy, Inc., under cooperative agreement with the National Science Foundation.} (IRAF) for detection and photometry of the sources.
On each wave-plate frame at $J$, $H$, and $K_s$, the fluxes of the detected sources were estimated by fitting Penny2-type point-spread function (PSF) to them.
The pixel coordinates of the sources were converted to celestial coordinates, based on the 2MASS All Sky Point Source Catalog \citep{skr06}.
The instrumental magnitudes were converted to 2MASS magnitude following \citet{kim16}.

The Stokes parameters $I$, $Q$, and $U$ of the sources are calculated using the following equations: $I$ $=$ ($I_0$+$I_{22.5}$+$I_{45}$+$I_{67.5}$)/2, $Q$ $=$ $I_0-I_{45}$, and $U$ = $I_{22.5}-I_{67.5}$, where $I_0$, $I_{22.5}$, $I_{45}$, and $I_{67.5}$ are the fluxes on each wave-plate angle frame (0$\arcdeg$, 45$\arcdeg$, 22\fdg5, and 67\fdg5).
We calculated the flux errors of the sources and derived their polarization uncertainties as described in \citet{kim16}.
The debiased degree of polarization, $P$, and the polarization angle, $\theta$, were calculated as described in \citet{kim11}.

\subsection{Separation of Foreground Sources}
\begin{figure*}[p]
\begin{center}
\includegraphics[scale=0.6]{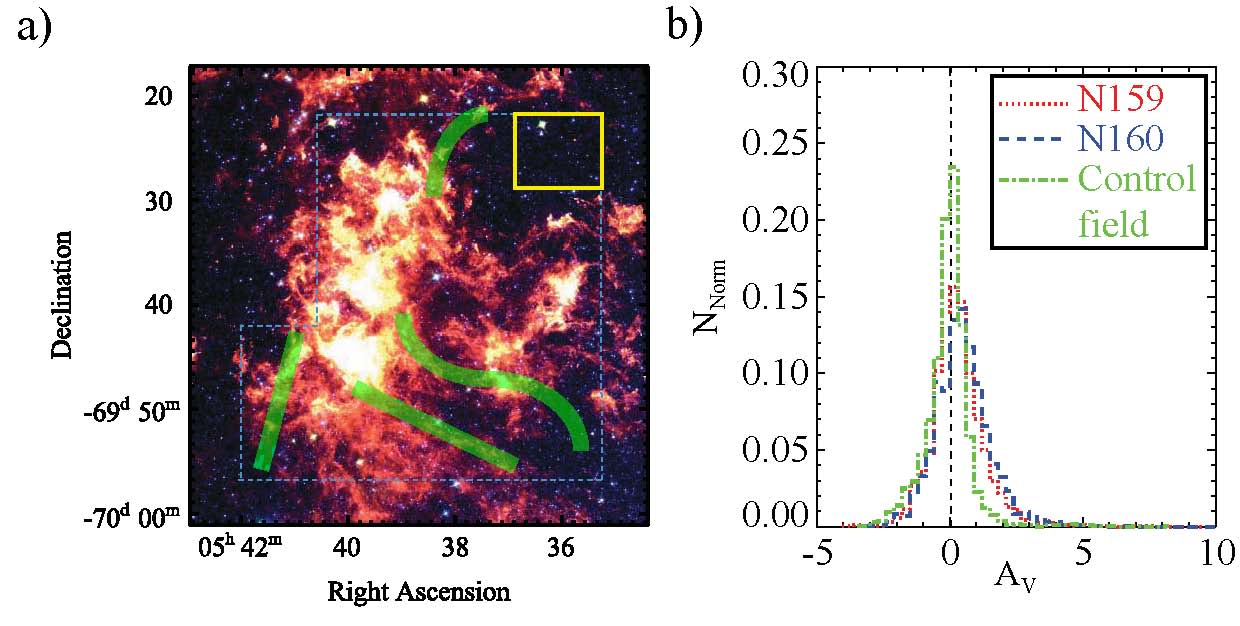}
\caption{a) Location of the extinction-free control field (yellow box) for the NICER technique. The Background color composite image describes dust emission features in the molecular ridge, using the IRAC data (3.6, 5.8, and 8$\mu$m) from the \textit{Spitzer} SAGE \citep{meix06}. Large-scale magnetic fields around the N159/N160 complex are shown by green-shaded curves, as seen in Figure 12 of \citet{kim16}. b) Normalized histograms of the visual extinction ($A_\mathrm{V}$) for the stars in the control (green dotted$-$dashed line), N159 (red dotted line), and N160 (blue dashed line) fields. The mean of $A_\mathrm{V}$ of the sources in the control field is 0.01 mag (vertical black dashed line), and the standard deviation is 0.47 mag.}\label{fig1}
\end{center}
\end{figure*}

The magnetically aligned dust grains generate the polarized light with E-vector direction parallel to the magnetic field, but only from the stars behind the field.
In addition, the light of the foreground stars in front of the ISM field is not affected by the magnetic field of our interest.
Therefore, the polarization of the foreground stars contaminates our results that reveal the magnetic fields of the N159/N160 complex.
We separated the foreground stars out from the detected sources using their visual extinction, $A_\mathrm{V}$, derived with the Near-Infrared Color Excess Revisited (NICER) technique \citep{lom01}.
The NICER technique is the revised version of that of \citet{lad94}, and it can measure the interstellar extinction from multi-band observations of molecular clouds.
We estimated visual extinctions of stars in two target fields and a control field with their near-infrared colors using the NICER algorithm\footnote{Using the C. Beaumont Interactive Data Language (IDL) implementation}.
The control field (yellow box in Figure~\ref{fig1}(a)) is a nearby extinction-free region centered at ($\alpha$, $\delta$)$_{2000}$ $=$ (5\textsuperscript{h}36\textsuperscript{m}08\textsuperscript{s}.3, -69$\arcdeg$26$\arcmin$00\farcs2).
The $JHK_s$ photometric data of the control field are from \citet{kim16}.

Figure~\ref{fig1}(b) shows the distribution of visual extinctions in the control and target fields.
The histogram for the control field peaks at 0.01 mag and with a standard deviation of 0.47 mag.
Based on these values, we regarded the sources with $A_\mathrm{V}$ $\leq$ 0.48 mag as the foreground stars in front of our target field.
A total of 1062 and 640 sources in the N159 and N160 fields, respectively, remained as the background stars for examining the polarimetric properties.

\section{Results}
\subsection{Highly Polarized Sources}
Highly polarized sources (HPSs) are mostly due to the intrinsic polarization by circumstellar dust around the stars themselves, not due to the dichroic absorption by foreground dust \citep{kan07}.
These sources should be separated from from our polarization sources to trace the magnetic fields that make interstellar dust align.

\begin{figure*}[p]
\begin{center}
\includegraphics[scale=0.4]{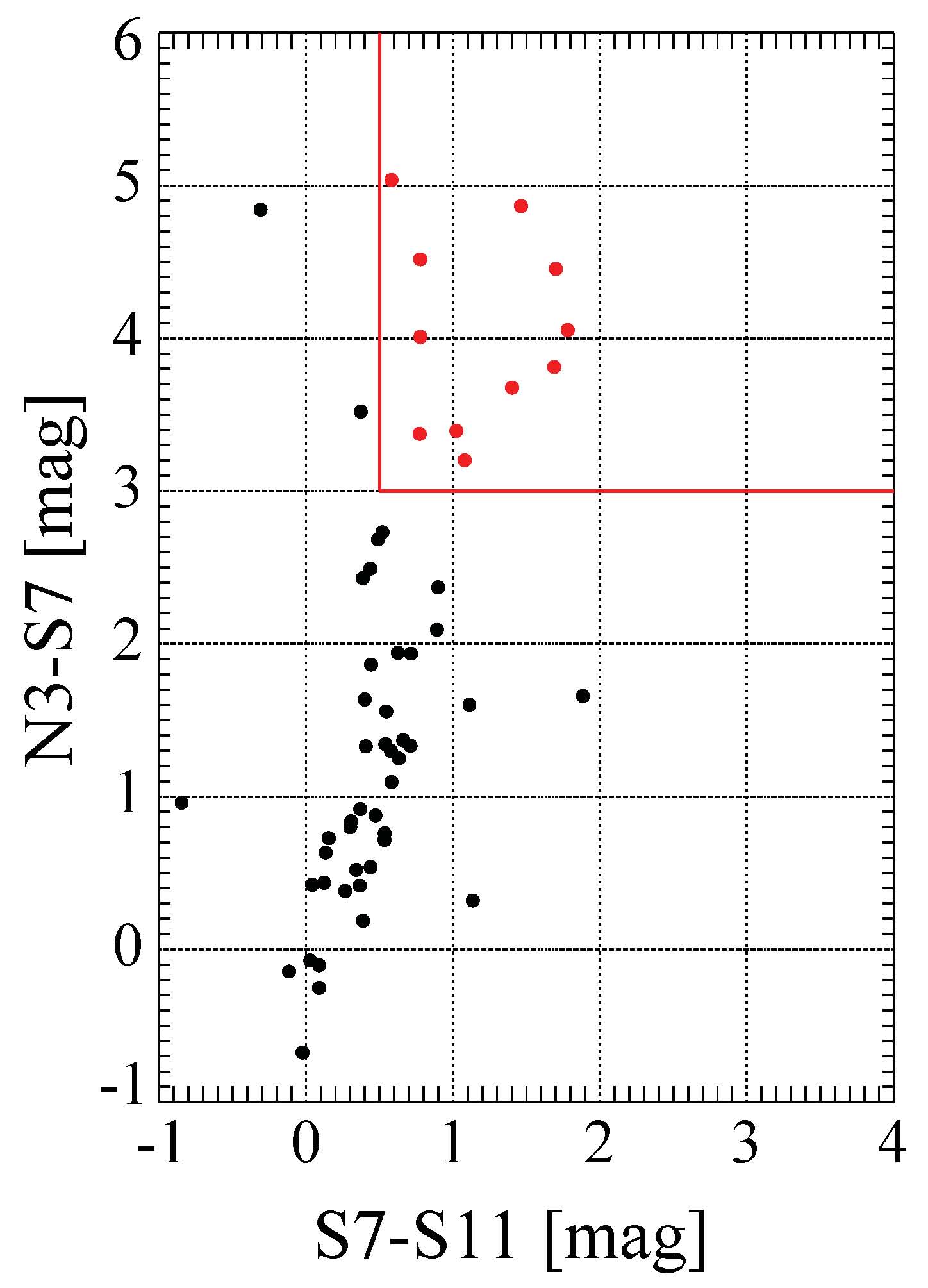}
\caption{Color$-$color (N3$-$S7 vs. S7$-$S11) diagram of the sources matched with those of \citet{kat12}. Most of the sources follow the dusty C-rich star sequence as described by \citet{kat12}. The sources inside red box (red circle) represent the possible YSOs with water ice absorption \citep{kat12}.}\label{fig2}
\end{center}
\end{figure*}

YSOs are one of the plausible sources of the intrinsic polarization, due to the dust grains in their circumstellar disk and envelope.
The N159/N160 complex contains quite a number of YSOs.
We matched our sources with the YSO candidates from \citet{che10} and \citet{car12}.
In addition, we used the catalog of the \textit{AKARI} Infrared Camera (IRC) survey in the LMC \citep{ita08,kat12} and classified new YSO candidates in this complex, using the color-color diagram (Figure~\ref{fig2}).
Most sources in the diagram follow the sequence of dusty C-rich stars denoted group CC3 in \citet{kat12}, but the sources redder than CC3 were found to be above [N3]$-$[S7] $\sim$ 3.0 mag.
\citet{shi08} mentioned that these redder sources are the YSO candidates with the water ice absorption at 3 $\mu$m.
We considered the nine redder sources to be new YSO candidates in the N159/N160 complex.
Two HEBs and one YSO from \citet{mar08} were also detected in our observations.

A total of 45 YSO candidates and 2 HEBs in previous works were found in our catalog.
Twelve of them show polarization degrees larger than 10 \%, and polarization signal-to-noise ratios (hereafter $P/\sigma_P$) greater than 3.
In Table~\ref{tab-1}, we compiled near-infrared photometric, polarimetric data, and information from other literature for the all 45 matched YSO candidates.

\begin{figure*}[p]
\begin{center}
\includegraphics[scale=0.55]{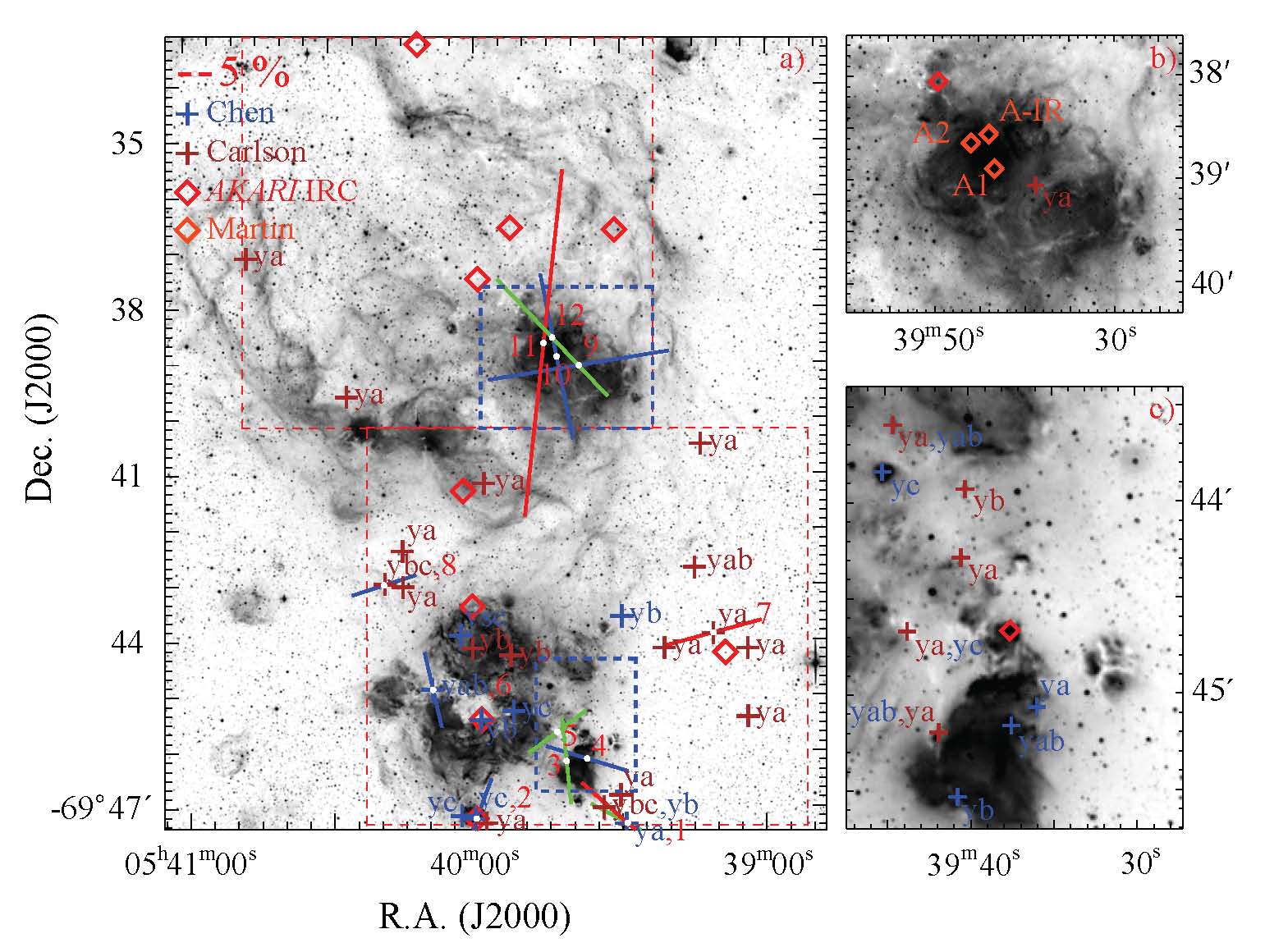}
\caption{a) Polarization vector map for the N159/N160 complex overlaid on the H$\alpha$ image from the WFI/ESO telescope. Highly polarized sources that are matched with the YSO candidates and have $P/\sigma_P$ $>$ 3 and $P$ $>$ 10 \% at each band are labeled by number. The color of lines denotes $J$ (blue), $H$ (green), and $K_s$ (red). The lengths of the vectors denote the degree of polarization, the scale of which is shown in the upper left corner. The YSO candidates from the literature are marked with blue plus symbols \citep{che10}, and brown plus symbols \citep{car12}. The evolutionary stages of the YSO candidates are labeled `ya', `yb', and `yc', indicating their stages from \textrm{I} to \textrm{III}, respectively, as in \citet{che10} and \citet{car12}. The label `ysc' indicates a young stellar cluster \citep{che10}, and the open orange diamonds are HEBs (A1 and A2) and a massive YSO (A-IR) \citep{mar08}. The open red diamonds represent the $AKARI$ IRC sources \citep{kat12} matched to our polarization sources. Panels (b) and (c) enlarge the regions inside the blue dashed boxes to clarify the massed positions of the YSO candidates in N160 and N159.}\label{fig3}
\end{center}
\end{figure*}

Figure~\ref{fig3} a) shows the polarization vectors of the 12 highly polarized sources and locations of the matched YSO candidates on the H$\alpha$ image from the WFI/ESO archive.
We excluded the highly polarized sources listed in Table~\ref{tab-1} in the analysis.
The Other 35 YSO candidates did not satisfy the polarization quality criterion of $P/\sigma_P$ $>$ 3 or showed the polarization pattern caused by the dichroic extinction due to interstellar dust.
A fraction of the intrinsic component might exist in their polarization, but the dominant fraction is probably the dichroic effect due to the aligned interstellar dust, which is strong enough to make the intrinsic component negligible.

\subsection{Estimation of the Polarization Accuracy}
In order to study the magnetic field in \textbf{the} N159/N160 complex, we set the selection criterion of $P/\sigma_P$ $>$ 3.
Most sources with this criterion showed their polarization degrees smaller than 10 \% \citep{kan07,nak07,san14}.

\begin{figure*}[p]
\begin{center}
\includegraphics[scale=0.6]{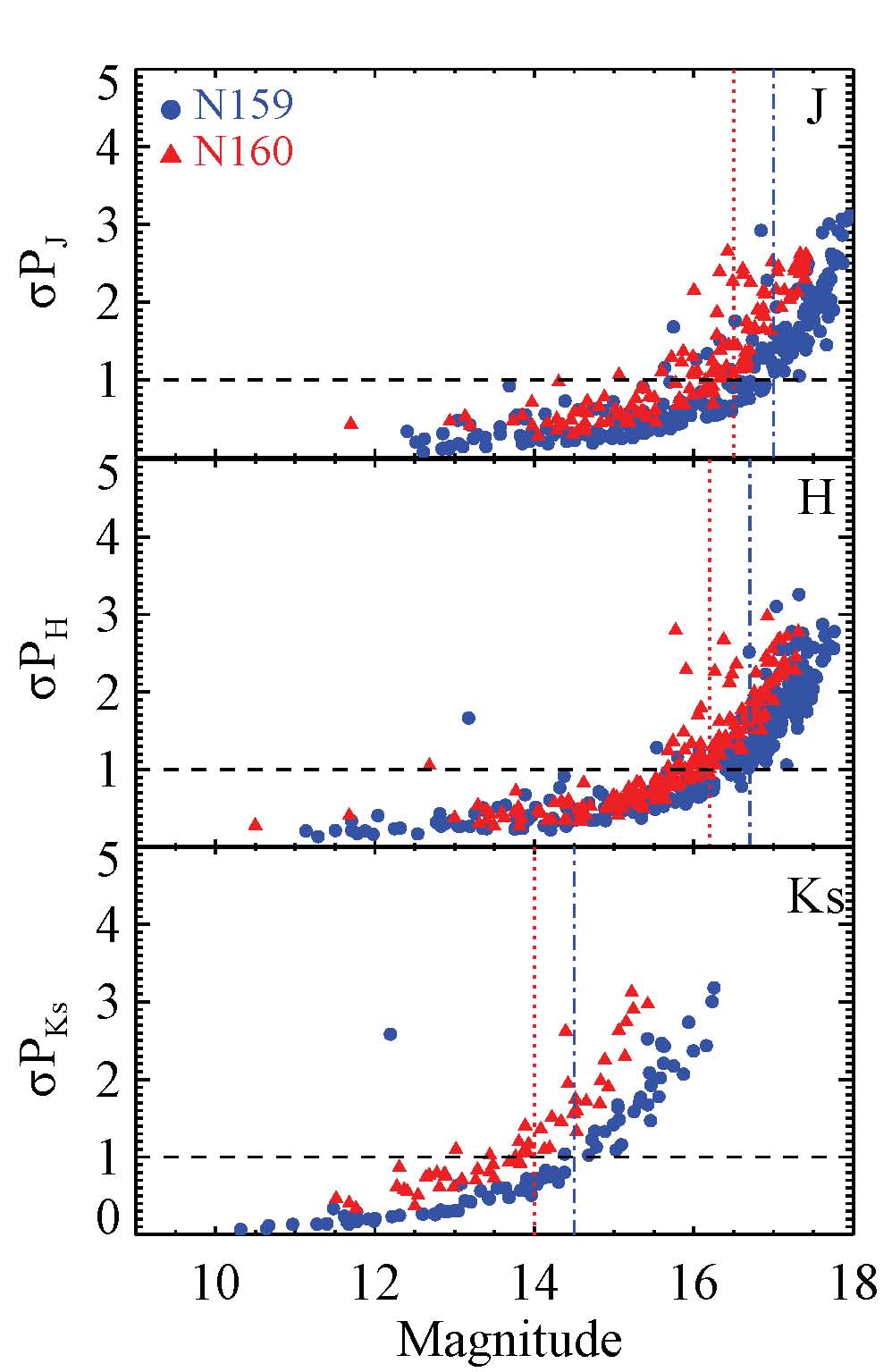}
\caption{Polarization uncertainty as a function of magnitude in the $J$ (top), $H$ (middle), and $K_S$ (bottom). Vertical dotted$-$dashed and dotted lines indicate the magnitude limits in N159 (red) and N160 (blue) that certify accurate polarization. The blue circles and red triangles indicate the sources in N159 and N160, respectively. Horizontal dashed lines indicate $\sigma_P$ $=$ 1 \%.}\label{fig4}
\end{center}
\end{figure*}

In general, fainter sources have larger polarization uncertainties than brighter ones.
Figure~\ref{fig4} shows the trend between polarization uncertainty and the magnitude of our sources.
We set the photometric limits at $J$, $H$, and $K_s$ to be 17.0, 16.7, and 14.5 mag for the N159 field and 16.5, 16.2, and 14.0 mag for the N160 field, respectively, so that stars brighter than these limits have a polarization uncertainty smaller than 1 \%.
We assume that the polarization of sources brighter than these magnitude limits are measured with sufficiently a high accuracy.
The criteria for selecting final catalog sources are given in Table~\ref{tab-2}.
Table~\ref{tab-3} lists the photometric and polarimetric data of the selected sources in the N159 and N160 fields.
The $A_\mathrm{V}$ values obtained using the NICER algorithm are also listed.

\begin{figure*}[p]
\begin{center}
\includegraphics[scale=0.8]{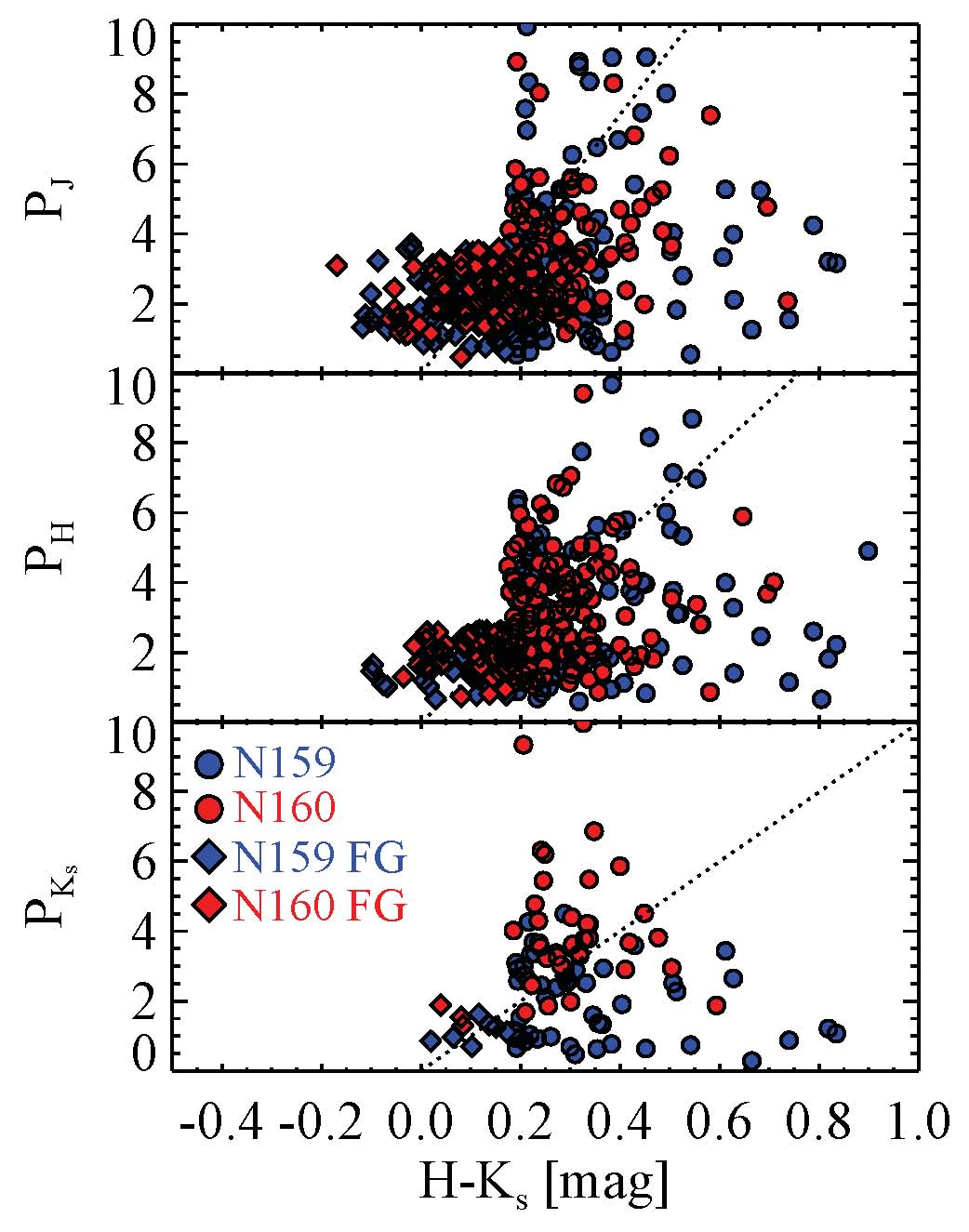}
\caption{Polarization degree vs. color ($H$-$K_s$) diagrams at $J$ (top), $H$ (middle), and $K_s$ (bottom) for the sources in the N159 (blue filled circle) and N160 (red filled circle) regions using the criterion of the qualified polarization. Dotted lines are empirical upper limits, $P_{\rm max}$ \citep{jon89}. Most sources have a polarization degree below this maximum polarization of interstellar origin. Polarization degrees of the foreground group (red and blue filled diamonds) in general exceed the limits of the interstellar polarization in N159 and N160.}\label{fig5}
\end{center}
\end{figure*}

In order to confirm the interstellar polarization of the catalog sources, we examined the efficiency of the interstellar polarization for the 252, 277, and 89 sources in the N159/N160 complex.
In general, interstellar polarization gives the upper limit of the polarization degree, depending on the near-infrared color \citep{jon89}.
Figure~\ref{fig5} shows that most of the sources have polarization degrees lower than the upper limit of the interstellar polarization at each band.
On the contrary, most of the foreground sources exceed the upper limits.
Hence, we conclude that the polarization of the sources is mainly caused by interstellar dust aligned due to the magnetic field in the N159/N160 complex.

\subsection{Wavelength Dependence of Polarization}
\citet{ser75} empirically modeled the wavelength dependence of polarization for the Milky Way.
The empirical relationship can be represented by a power law of $P_{\lambda}$ $\propto$ $\lambda^{-\beta}$, with $\beta$ of 1.6$-$2.0 at the near-infrared wavelengths from 1.25 to 2.2 $\mu$m \citep{mar90,nagt90,mar92}.
In the case of the LMC, however, the relation has a shallower slope than that of the Milky Way, with $\beta$ $=$ 0.9 \citep{nak07,kim16}.

\begin{figure*}[p]
\begin{center}
\includegraphics[scale=0.7]{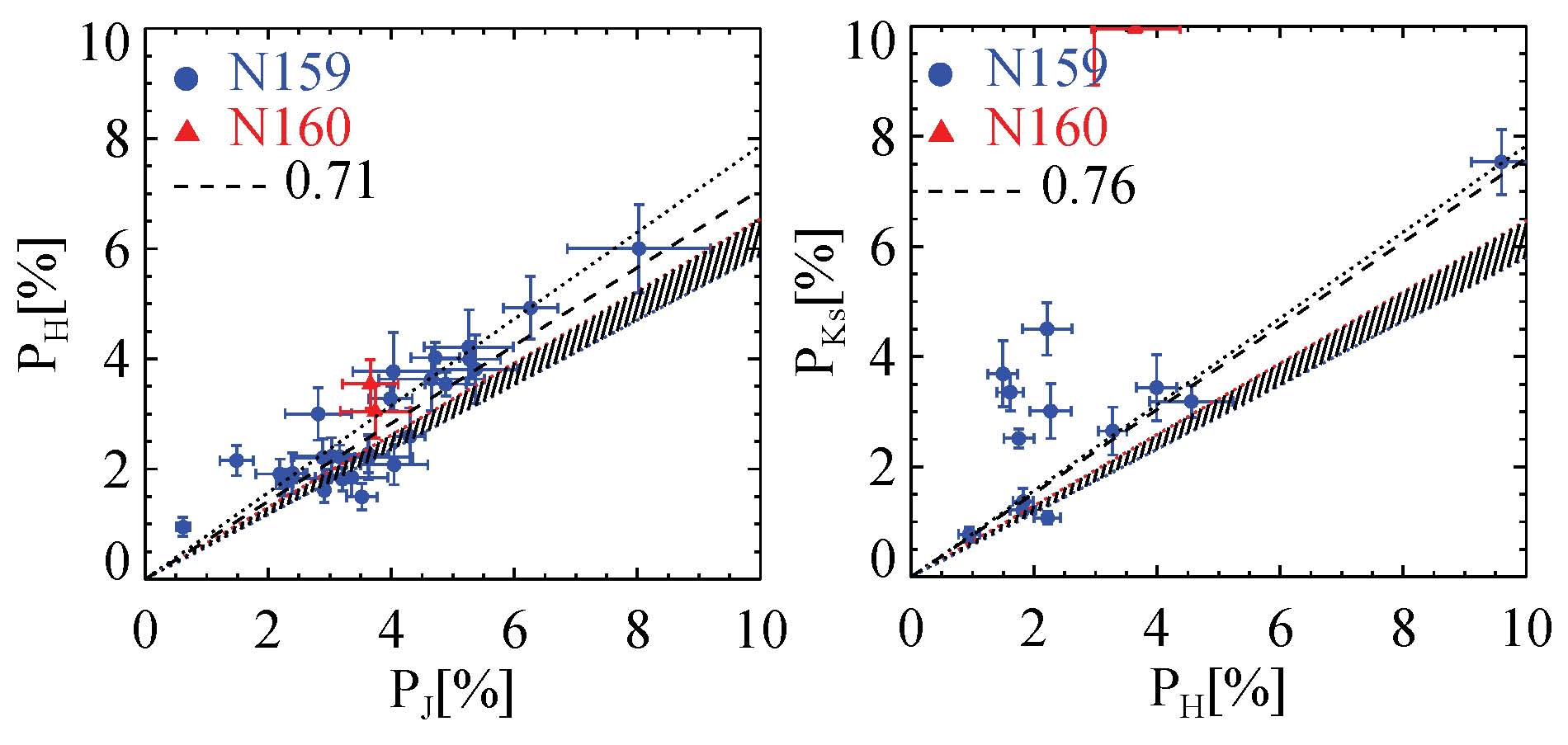}
\caption{Plots of $P_{J}$ vs. $P_{H}$ (left) and $P_{H}$ vs. $P_{K_{s}}$ (right) for the polarization sources with $P/\sigma_P$ $>$ 5 and $\sigma_P$ $\leq$ 1 in N159 (blue filled circle) and N160 (red filled triangle). The dashed lines denote the best-fit slopes for the combined sources of N159 and N160. The hatched area shows the range of the wavelength dependence from previous studies \citep{mar90,nagt90,mar92} for the Milky Way, and the dotted lines are the best-fit slopes from \citet{nak07}.}\label{fig6}
\end{center}
\end{figure*}

Figure~\ref{fig6} plots our selected sources with $P/\sigma_P$ $>$ 5 and $\sigma_P$ $\leq$ 1 \% to show the wavelength dependence of the polarization degrees.
The best-fit slopes of $P_H$/$P_J$ $=$ 0.71 and $P_{K_s}$/$P_H$ $=$ 0.76 indicate that the interstellar polarization of the N159/N160 complex also follows the power-law index of $\beta$ $=$ 0.9, rather than the case of the Milky Way.

\subsection{Linear Polarization Vector Map}
\begin{figure*}[p]
\begin{center}
\includegraphics[scale=0.7]{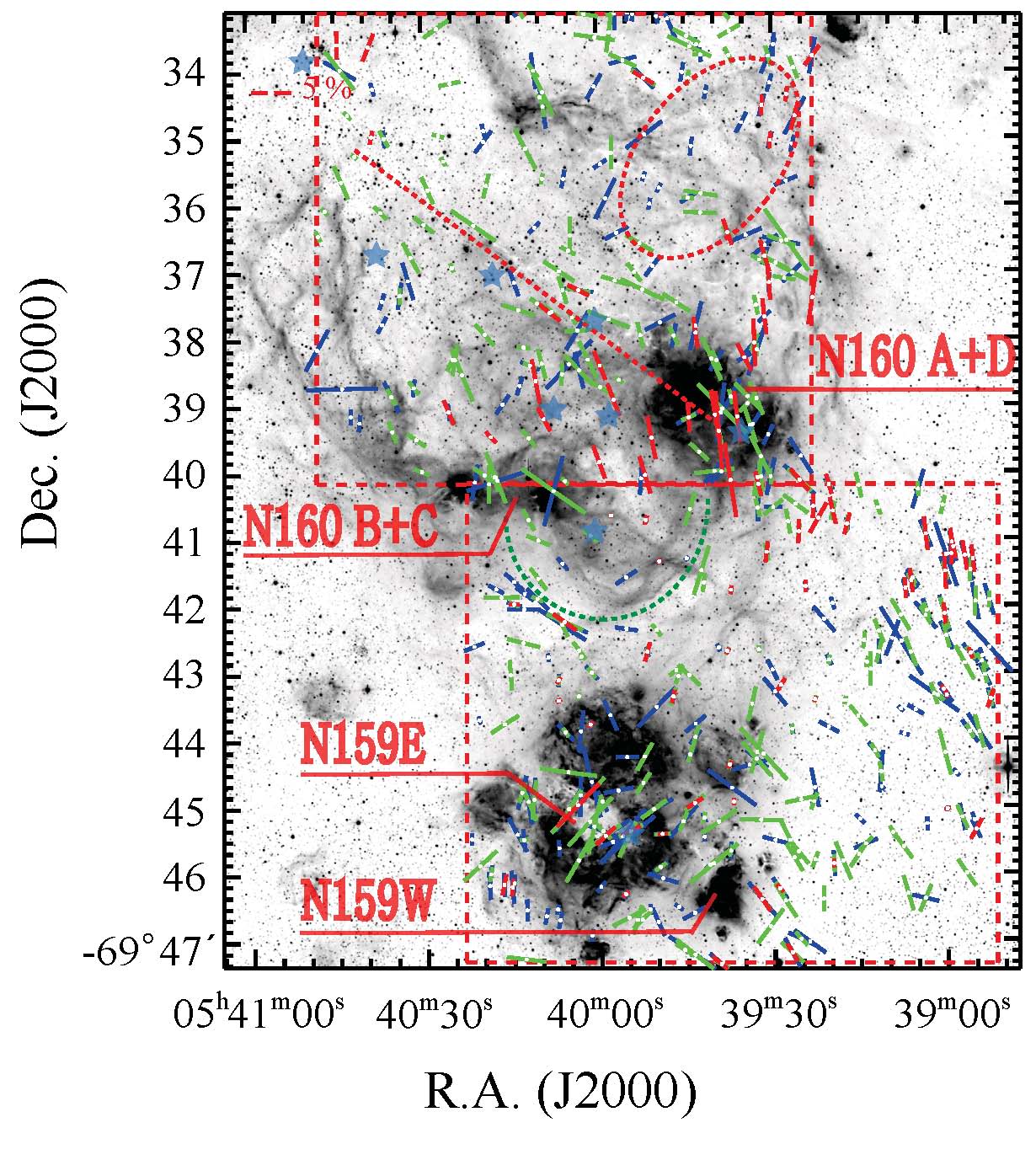}
\caption{Polarization vector map of the final catalog of sources for the N159/N160 complex overlaid on the H$\alpha$ image from the WFI/ESO telescope. \ion{H}{2} regions in N160 are denoted by N160 B+C and N160 A+D. Two GMCs in N159 are also denoted by N159E and N159W. Stars in our catalog are displayed with blue ($J$), green ($H$), and red ($K_s$) bars. The lengths of the bars are proportional to the degree of polarization with the scale shown in the upper left corner. The fields of observations in this study are marked with the red dashed line boxes. Interesting features in the H$\alpha$ emission of N160 are marked with colored dashed lines: probable axis of the expanding shell with the red straight line, northwestward expansion with the red ellipse, and southward U-shaped expansion with the green curved line. Locations of OB clusters \citep{nak05} in N160 are displayed with blue star symbols.}\label{fig7}
\end{center}
\end{figure*}

We can assume that the direction of a polarization vector is parallel to the sky-projected component of the magnetic field.
In Figure~\ref{fig7}, we draw the polarization vectors of our best-quality sources on the H$\alpha$ image to map the magnetic field structure of the N159/N160 complex.
The magnetic field structure shows different trends between N159 and N160.

In the N160 region, polarization vectors are roughly aligned with the axis of the open H$\alpha$ shell.
However, the vectors in the northwestern region (red dotted ellipse in Figure~\ref{fig7}) clearly follow a different pattern, when compared with those inside of the H$\alpha$ shell.
The vectors in the south of N160 B+C and N160 A+D show a U-shaped distribution (green dotted curve in Figure~\ref{fig7}), which is similar to the H$\alpha$ feature in this region.
The magnetic field structure inferred from the distribution of polarization vectors in the N160 region supports the expanding shell in the region.

N159 shows a complex magnetic field structure associated with H$\alpha$ emission.
Polarization vectors in N159E are distributed around the lower part, while those in N159W are on the outskirts of the H$\alpha$-emitting cloud.
The western side of N159 has no significant H$\alpha$ emission, and the polarization vectors in this region are mostly aligned with the polarization angle of about 30$\arcdeg$.

\begin{figure*}[p]
\begin{center}
\includegraphics[scale=0.8]{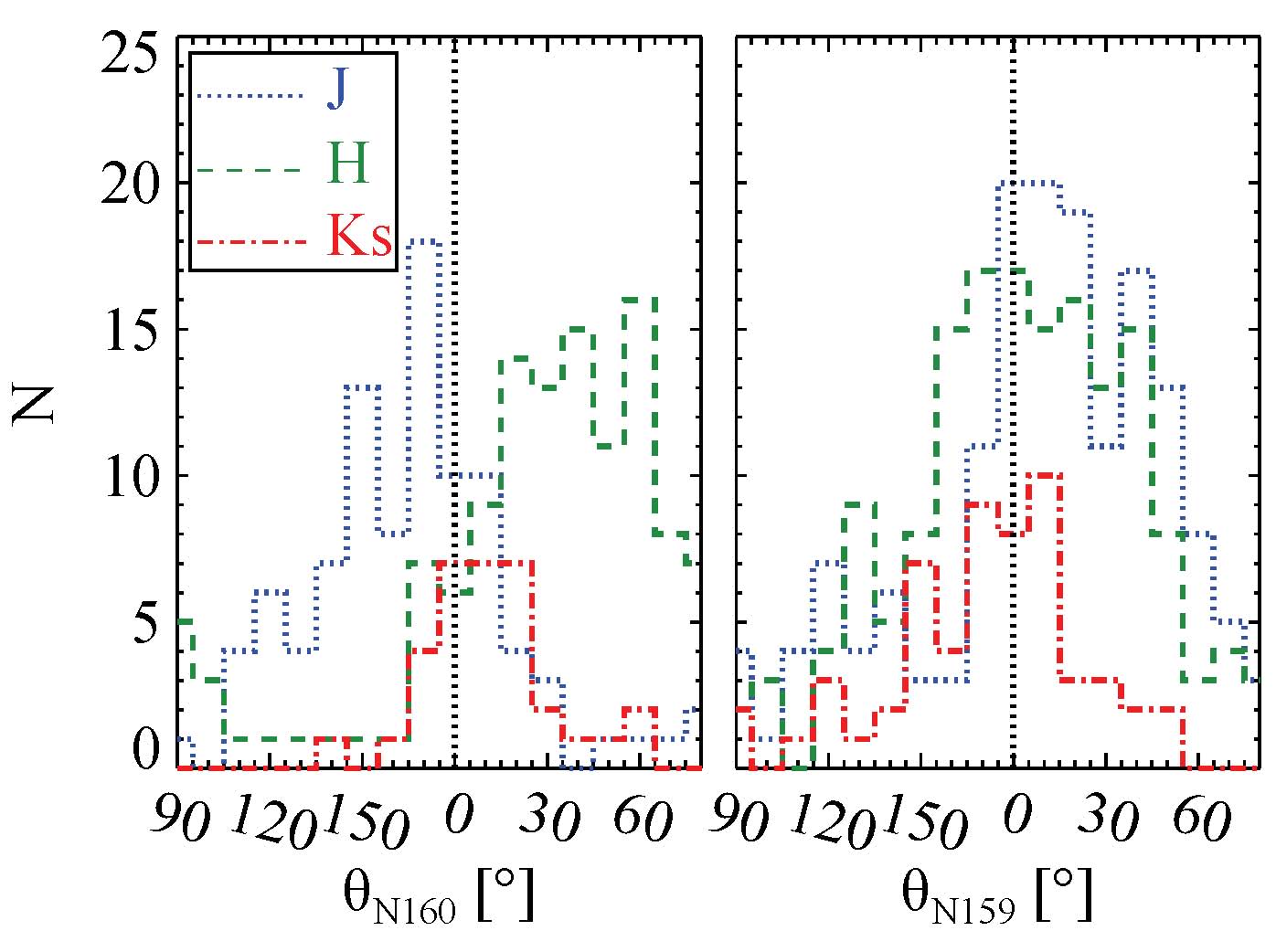}
\caption{Histograms of polarization angles in N160 (left) and N159 (right) at $J$ (blue dotted line), $H$ (green dashed line), and $K_s$ (red dotted$-$dashed line). The width of each bin is fixed to 10$\arcdeg$.}\label{fig8}
\end{center}
\end{figure*}

The peaks in the distribution of the polarization angles of each region in Figure~\ref{fig8} are good indicators of the major directions of the magnetic fields.
Those directions are also appeared in the polarization vector map (Figure~\ref{fig7}).
In the case of N160, one of the two peaks, around 30$\arcdeg$, is related to the magnetic field inside the H$\alpha$ shell, while the other, around 170$\arcdeg$, is related to the field on the northwestern part.
Histograms for N159 have a single peak around 10$\arcdeg$$-$30$\arcdeg$ and this is mostly caused by the magnetic field on the western side of the region.
The polarized sources in N159E and N159W have various polarization angles and are responsible for the relatively large dispersion of the histogram compared to that for N160.

\section{Discussion}

\subsection{Magnetic Field Structures with Molecular Cloud Emissions}
\begin{figure*}[p]
\begin{center}
\includegraphics[scale=0.7]{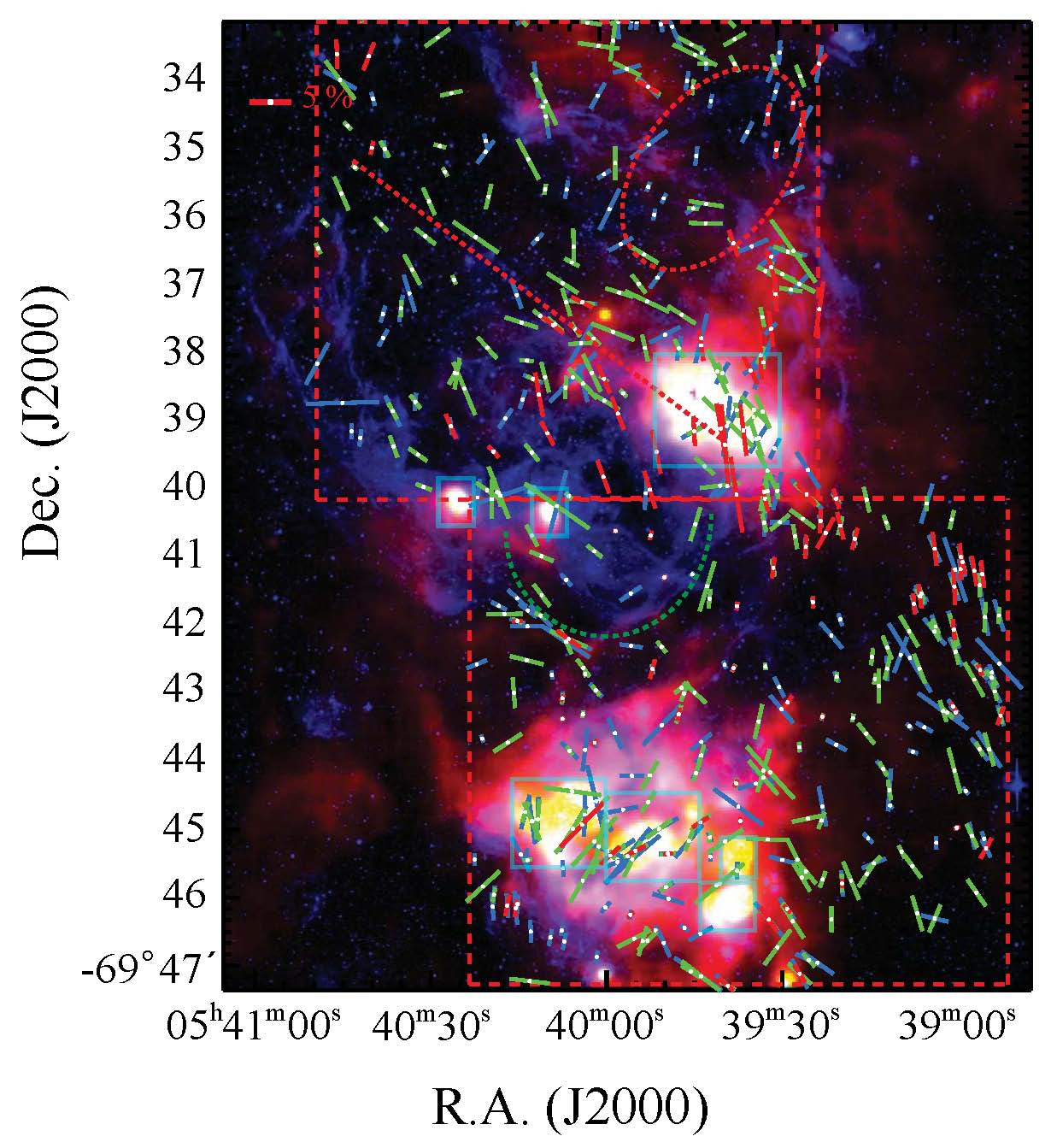}
\caption{Distribution of polarization vectors for the N159/N160 complex overlaid on the color composite image of \textit{Herschel} PACS 100 $\mu$m (red), \textit{Spitzer} IRAC 24 $\mu$m (green), and H$\alpha$ emission of the WFI/ESO telescope (blue). The dust emissions at 24 and 100 $\mu$m are denoted by cyan boxes. See the caption of Figure~\ref{fig7} for the meanings of the lines and symbols.}\label{fig9}
\end{center}
\end{figure*}

The magnetic field structures in star-forming regions are constrained by the environment (gas and dust).
To reveal the nature of the magnetic field structures in the N159/N160 complex, we have compared our data with the dust and gas emissions.
To trace the emission features of dust and ionized gas in the N159/N160 complex, the IRAC 24 $\mu$m data from the \textit{Spitzer} SAGE \citep{meix06} and the \textit{Herschel} 100 $\mu$m data from the HERITAGE project \citep{meix10} were combined with the vector map on the H$\alpha$ image(Figure~\ref{fig9}).

As seen in Figure~\ref{fig9}, most of the H$\alpha$ shell structures in N160 encounter the boundary of dust emission at 100 $\mu$m.
This trend is more clearly seen at the southern shell structure rather than the northern one.
The red dotted ellipse region of the northern shell shows a leak feature of the H$\alpha$ emission extending to the dust cavity, and the polarization vectors in this region are well-aligned in the direction of the cavity.
The southern regions of N160 B+C and N160 A+D show the U-shaped feature of the polarization vectors at the edge of strong H$\alpha$ and 100 $\mu$m emissions.
In addition, the magnetic field structures inside the shell structure are distributed along the direction from northeast to southwest, which is consistent with the triggered direction of the sequential cluster formation suggested by \citet{nak05}.
We also found that some polarization vectors are seen along the shell structure like an expanding shell.

\begin{figure*}[p]
\begin{center}
\includegraphics[scale=1.5]{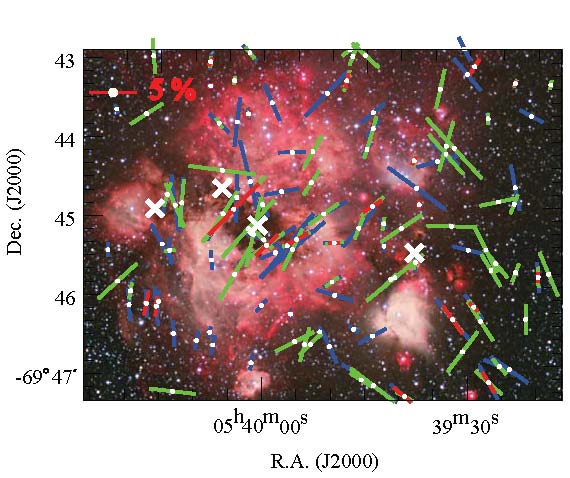}
\caption{Locations of the integrated intensity peaks of $^{12}$CO($J$ $=$ 3$-$2) (white X symbol). The polarization vectors in the N159 region at $J$ (blue), $H$ (green), and $K_s$ (red) are displayed on this composite $B$, $V$, and H$\alpha$ image.}\label{fig10}
\end{center}
\end{figure*}

Polarization vectors in the N159 region are mostly found near \ion{H}{2} regions showing the dust emissions at 24 $\mu$m and 100 $\mu$m (yellow-composite feature in Figure~\ref{fig9}).
In order to understand the complex structure of the magnetic fields in N159, we compared our vector map with the CO gas distribution.
\citet{miz10} observed the $^{12}$CO($J$ $=$ 4$-$3) and $^{12}$CO($J$ $=$ 3$-$2) rotational lines in N159 with the NANTEN2 \citep{fuk08} and ASTE \citep{eza04} submillimeter telescopes, respectively.
We compared the integrated intensity map of $^{12}$CO($J$ $=$ 3$-$2) \citep{miz10} with our polarization vector map on the H$\alpha$ image in Figure~\ref{fig10}.
Strong concentrations of the CO distribution (white X symbols in Figure~\ref{fig10}) are located at the regions where H$\alpha$ emission is obscured by the high extinction of the molecular gas.
In the case of N159W, the magnetic fields appear around the 24 $\mu$m emission and CO peak.
The configuration of the complex magnetic field structure in N159 appears to have been affected by the expansion and evolution of the \ion{H}{2} regions in the central N159 and the southern N159W.
\citet{jon05}, \citet{che10}, and \citet{ber16} proposed expansions of the \ion{H}{2} regions in N159E and N159W.
They also reported that the locations of several compact \ion{H}{2} regions and young stellar clusters in N159 are likely to be linked with the molecular gas and are mainly distributed on the edge of the ionized \ion{H}{2} regions in N159E and N159W.

The western region of N159 does not show any interesting features of dust and gas emissions, while the magnetic fields are in a uniform direction, as shown by the polarization vectors at this region.
It indicates that this region was not affected by the star-forming activities in the N159/N160 complex and that a larger scale magnetic field was already formed toward this region before the beginning of the star formation in the nascent molecular cloud in N159 and N160.

In conclusion, the polarization vector map with dust and gas emission features indicates that the magnetic field structures in N159 and N160 were affected by different star-forming activities in each region.
These different patterns of the magnetic field support the suggestion that N159 and N160 are in the different evolutionary stages based on the large-scale sequential star formation.

\subsection{Formation Scenario of the N159/N160 Complex and Magnetic Field Structure}
The spatial distributions of the Herbig Ae/Be and OB clusters in the N159/N160 complex at the near-infrared bands were revealed by \citet{nak05}.
Using the spatial correlations between these young stellar clusters and gaseous components, they suggested a large-scale cluster formation scenario that occurred sequentially from north to south.
Possible starting points of the formation are the supergiant shell LMC 2 and a superbubble located northeast and east of the complex, respectively.
However, the dominant magnetic field direction in N160 shows the pattern along the northeast to southwest, which is in the same direction to the LMC 2 rather than that to the superbubble.
The similar direction between the magnetic field and the sequential formation shows that the LMC 2 is a plausible trigger for this large-scale cluster formation.

\begin{figure*}[p]
\begin{center}
\includegraphics[scale=0.28]{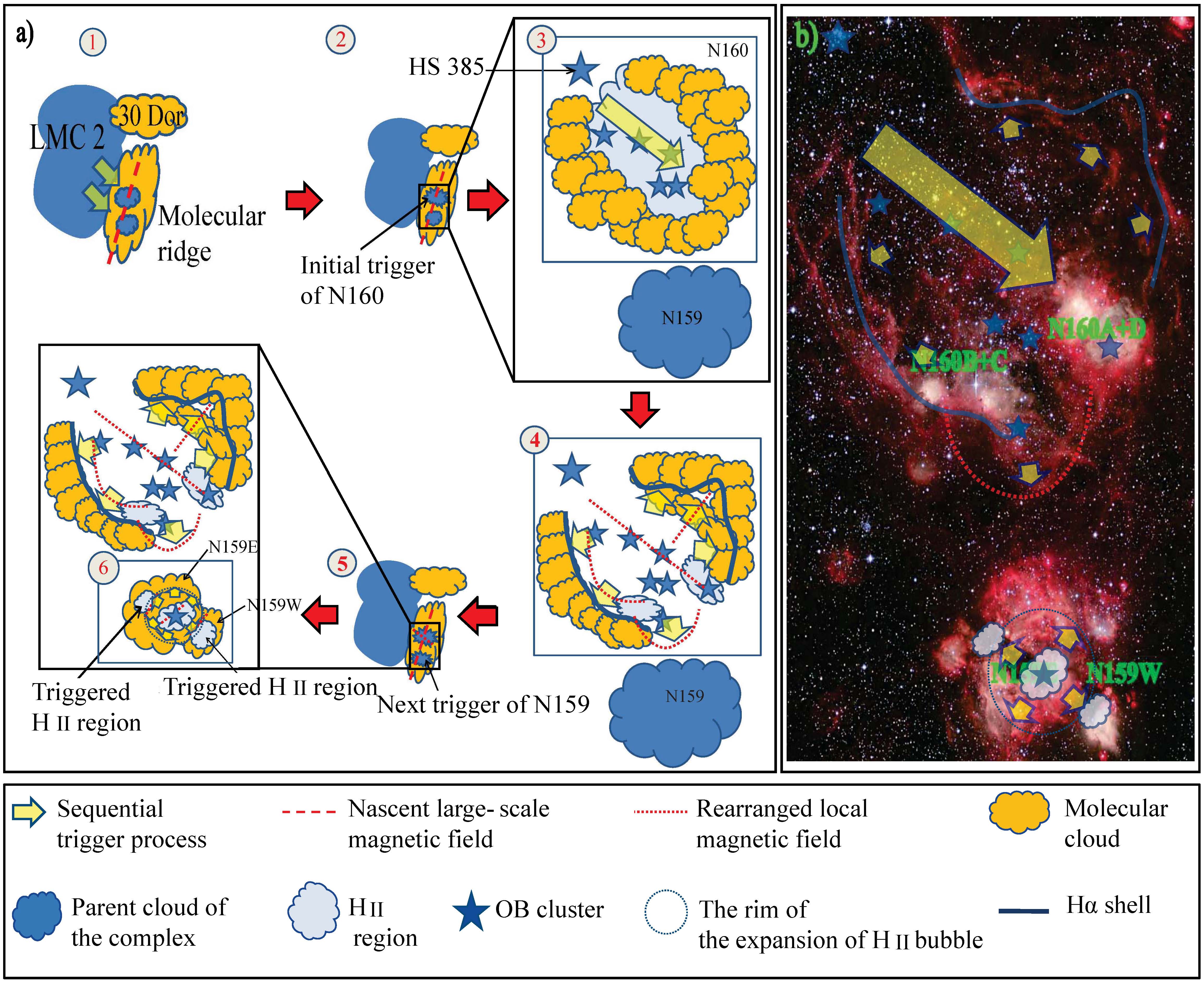}
\caption{a) Illustration of the sequential formation process of the N159/N160 complex and the magnetic field structures. b) the N159/N160 star-forming complex on the RGB (Red: H$\alpha$, Green: $V$, and Blue: $B$) composite image from the WFI/ESO telescope. \ion{H}{2} regions in N160 are denoted as N160 B+C and N160 A+D. The shell structure of N160 is clearly seen through its H$\alpha$ emission feature (blue curve). Two GMCs in N159 are also denoted as N159E and N159W.}\label{fig11}
\end{center}
\end{figure*}

Figure~\ref{fig11} (a) shows an illustration of the sequential formation process of the N159/N160 complex and the magnetic field structures at each scenario step.
The RGB (Red: H$\alpha$, Green: $V$, and Blue: $B$) composite image of the complex is also displayed to help in understanding each illustrated feature (Figure~\ref{fig11} (b)).
LMC 2 (blue-colored amorphous feature), the molecular clouds (orange-colored cloudy feature) of the 30 Doradus complex, and the molecular ridge are shown in Step 1, based on the large-scale H$\alpha$ map of Figure 5(a) in \citet{poi01}.
\citet{kim99} classified \ion{H}{1} shell structures in the LMC by using their shell size, expansion pattern, morphological structure, and associated H$\alpha$ emission.
The LMC 2 was classified as an expanding supergiant shell.
At Step 1, the interaction between the expansion of LMC 2 and the parent cloud of the complex might influence the initial star formation at N160.
As we showed in our previous paper \citep{kim16}, the boundary between the western border of LMC 2 and the large-scale pattern of magnetic fields (red dashed line) is well-aligned in the eastern side of the molecular ridge.
This may support the process of the expansion of LMC 2 and its triggering of star formation.

At Step 2, the OB star formation begins in the parent cloud of N160.
The first-formed cluster, HS 385, is located at the northeastern region of N160, as denoted by a big blue star symbol in Step 3.
We suggest that a shock wave driven by stellar radiation and stellar winds from HS 385 sequentially triggered the formation of OB stars through this parent cloud (yellow arrow in Step 3).
Actually, most of the OB clusters in the N160 region are detected inside of the shell, as denoted by blue-colored stars in Figure~\ref{fig7} and in this illustration.
In addition, their age distribution from \citet{bic96} indicated that HS 385 is older than other OB clusters in the N160 region, and the southwestern region of N160 contains the Herbig Ae/Be clusters, with their ages less than 3 Myr \citep{nak05}.

Step 4 illustrates the expansion of \ion{H}{2} regions energized by the OB clusters within the shell, which passed through the south of N160 B+C and N160 A+D, and toward the GMCs of N159.
The dominant direction of the magnetic field structure at the interior of the shell (red dotted line in Step 4) is coincided with that of the sequential formation process.
The magnetic fields near the southern shell trace its boundary as shown in Figure~\ref{fig9}.
The U-shaped magnetic field structure in the south of N160 B+C and N160 A+D is bent toward the central region of N159.
These features indicate that magnetic fields are associated with the expansion of the \ion{H}{2} region in N160.
It also implies that the path is propagated sequentially from N160 to the nearby GMC of N159.
A similar propagation process is also suggested in the Scorpius$-$Centaurus association by \citet{pre99,pre07}.
They proposed a scenario for the sequentially triggered star formation and dissipation of molecular clouds in three subgroups of Scorpius-Centaurus association, by stellar winds and ionizing radiation from the OB stars therein.

Steps 5 and 6 illustrate the triggered star formation and the related magnetic fields in N159.
According to \citet{jon05}, formation of OB stars at the center of N159 pushed \ion{H}{2} regions out to the surrounding molecular cloud, and the subsequent star formation then triggered at the rim of the expansion bubble (dotted circle in Step 6).
The distribution of star formation rates from \citet{gal13} supports this triggering process.
The magnetic fields in the central part of N159 are distribute on the \ion{H}{2} region that is considered to be the initial trigger site for N159.
Subsequent triggering processes can also be traced by the magnetic fields, following the boundary of the triggered \ion{H}{2} regions (sky-colored clouds in Step 6) located around the rim of the expansion bubble.
We suggest that the dynamical star-forming activities and expansion of the \ion{H}{2} bubble disturbed and rearranged the nascent magnetic field structure.
Uniformly aligned magnetic fields at the southeast and the west of N159, which do not show any star-forming activities and \ion{H}{2} regions, support the change in the magnetic field because of these effects.

\section{Summary}
We conducted near-infrared polarimetry for the stars in the N159/N160 star-forming complex in the LMC.
We applied the NICER algorithm to select background stars for the study of the magnetic field in our observation field.
The 36 YSO candidates and 2 HEBs were matched, and 12 of them were the sources that are highly polarized, likely by the dust in the circumstellar envelope rather than by the interstellar dust.
We excluded them from the final catalog to study only the interstellar polarization in the N159/N160 complex.
We newly verified nine YSO candidates using the color-color diagram for the sources matched with the $AKARI$ IRC LMC catalog.
The 252, 277, and 89 sources in the N159 and N160 fields with sufficiently good polarization qualities represent dichroic polarization from the interstellar dust in this complex.
As other studies \citep{nak07,kim16} have suggested, wavelength dependence of polarization turned out to be weaker here than in the case of the Milky Way.

We visualized the structure of the magnetic fields in the N159 and N160 fields, using a polarization vector map projected on the H$\alpha$ image.
Polarization vectors showed a complex distribution of the magnetic fields, indicating the interaction between star-forming activities and surrounding interstellar medium.
The comparison between the polarization vectors and molecular cloud emissions showed that the magnetic fields are resulted from the different formation histories of N159 and N160.
From the distribution of the magnetic field structures, we suggested the plausible scenario that the star formation is sequentially triggered from N160 to N159.
We also proposed that ionizing radiation from OB clusters and the expanding \ion{H}{2} bubbles in this complex probably affected the nascent magnetic field structure.

\acknowledgments
This work was supported by the National Research Foundation of Korea (NRF) grant No. 2008-0060544, funded by the Korean government (MSIP).
We would like to thank Prof. Yasushi Nakajima for kindly providing comments that improved this paper.
This paper uses observations performed at the South African Astronomical Observatory.
This publication makes use of data products from the Two Micron All Sky Survey and observations with AKARI. The Two Micron All Sky Survey is a joint project of the University of Massachusetts and the Infrared Processing and Analysis Center/California Institute of Technology, funded by the National Aeronautics and Space Administration and the National Science Foundation. The AKARI is a JAXA project with the participation of ESA.
This research has made use of the NASA/ IPAC Infrared Science Archive, which is operated by the Jet Propulsion Laboratory, California Institute of Technology, under contract with the National Aeronautics and Space Administration.

{}

\begin{deluxetable}{ccccccccccccccccccccccccccccc}
\tabletypesize{\tiny}
\rotate
\setlength{\tabcolsep}{0.02in}
\tablewidth{0pt}
\tablecaption{Photometric and polarimetric catalog of YSOs and HPSs the N159/N160 field \label{tab-1}}
\tablehead{
\multicolumn{2}{c}{Position} & \colhead{} & \multicolumn{6}{c}{Magnitude} & \colhead{} & \multicolumn{13}{c}{Polarization Properties} & \colhead{} & \colhead{} & \colhead{} & \colhead{HPS} & \multicolumn{2}{c}{AKARI color\tablenotemark{c}}\\

\cline{1-2} \cline{4-9} \cline{11-23} \cline{28-29}\\

\multicolumn{1}{c}{${{\alpha_{\circ\rm J2000.0}}}$} & \multicolumn{1}{c}{${\delta_{\circ\rm J2000.0}}$} & \colhead{} &
\multicolumn{2}{c}{$J$} & \multicolumn{2}{c}{$H$} & \multicolumn{2}{c}{$K_{s}$} & \colhead{} &
\multicolumn{2}{c}{$P_{J}$} & \multicolumn{2}{c}{$P_{H}$} & \multicolumn{2}{c}{$P_{K_{s}}$} & \multicolumn{1}{c}{} &
\multicolumn{2}{c}{$\theta_{J}$} & \multicolumn{2}{c}{$\theta_{H}$} & \multicolumn{2}{c}{$\theta_{K_{s}}$} & \colhead{Av} & \colhead{References} & \colhead{Type\tablenotemark{a}} & \colhead{No.\tablenotemark{b}} & \multicolumn{1}{c}{S7-S11} & \multicolumn{1}{c}{N3-S7}\\

\multicolumn{1}{c}{} & \multicolumn{1}{c}{} & \colhead{} &
\multicolumn{2}{c}{(mag)} & \multicolumn{2}{c}{(mag)} & \multicolumn{2}{c}{(mag)} & \colhead{} &
\multicolumn{2}{c}{(\%)} & \multicolumn{2}{c}{(\%)} & \multicolumn{2}{c}{(\%)} & \multicolumn{1}{c}{} &
\multicolumn{2}{c}{($\arcdeg$)} & \multicolumn{2}{c}{($\arcdeg$)} & \multicolumn{2}{c}{($\arcdeg$)} & \colhead{(mag)} & \colhead{} & \colhead{} & \colhead{} &
\multicolumn{1}{c}{(mag)} & \multicolumn{1}{c}{(mag)}\\

\colhead{(1)}&\colhead{(2)}&\colhead{}&
\colhead{(3)}&\colhead{(4)}&\colhead{(5)}&\colhead{(6)}&\colhead{(7)}&\colhead{(8)}&\colhead{}&
\colhead{(9)}&\colhead{(10)}&\colhead{(11)}&\colhead{(12)}&\colhead{(13)}&\colhead{(14)}&\colhead{}&
\colhead{(15)}&\colhead{(16)}&\colhead{(17)}&\colhead{(18)}&\colhead{(19)}&\colhead{(20)}&
\colhead{(21)}&\colhead{(22)}&\colhead{(23)}&\colhead{(24)}&\colhead{(25)}&\colhead{(26)}
}
\startdata
5 39 29.02 & -69 47 18.69 && 16.763 & 0.047 & 15.791 & 0.066 & 14.576 & 0.058 &&  3.87 & 3.86 & 12.57 & 3.41 & 18.49 & 3.73 &&  14.25 & 20.20 &  62.04 &  7.48 & 48.67 & 5.65 & 7.01 & 1 & ya & 1 & $\cdots$ & $\cdots$ \\
5 40  0.48 & -69 47 13.02 && 14.502 & 0.059 & 13.583 & 0.043 & 12.395 & 0.028 && 12.93 & 3.28 &  3.06 & 2.68 & $\cdots$ & $\cdots$ && 159.60 &  7.04 &  26.79 & 18.84 & $\cdots$ & $\cdots$ & 6.80 & 1,3 & yc,CC5 & 2 & 1.40 & 3.68 \\
5 40  3.40 & -69 47 10.02 && 15.461 & 0.007 & 14.849 & 0.006 & 14.707 & 0.013 &&  1.72 & 0.51 & $\cdots$ & $\cdots$ & $\cdots$ & $\cdots$ && 163.49 &  8.09 & $\cdots$ & $\cdots$ & $\cdots$ & $\cdots$ & 0.24 & 1 & yc & $\cdots$ & $\cdots$ & $\cdots$ \\
5 39 33.62 & -69 47  1.14 && 16.726 & 0.028 & 16.124 & 0.026 & 15.505 & 0.019 &&  3.13 & 1.93 & $\cdots$ & $\cdots$ &  2.40 & 2.17 && 143.59 & 15.05 & $\cdots$ & $\cdots$ & 37.1 & 19.17 & 3.23 & 1,2 & yb,ybc & $\cdots$ & $\cdots$ & $\cdots$ \\
5 39 40.60 & -69 46 30.43 && 16.861 & 0.055 & 16.278 & 0.063 & 15.546 & 0.060 &&  8.59 & 3.63 &  7.88 & 3.79 & $\cdots$ & $\cdots$ &&  27.48 & 11.14 & 145.28 & 12.39 & $\cdots$ & $\cdots$ & 3.86 & 1 & yb & $\cdots$ & $\cdots$ & $\cdots$ \\
5 39 41.68 & -69 46 11.31 && 16.412 & 0.036 & 14.416 & 0.056 & 12.193 & 0.045 && $\cdots$ & $\cdots$ & 13.03 & 3.48 &  9.61 & 2.58 && $\cdots$ & $\cdots$ &   5.98 &  7.38 & 7.14 & 7.43 & 13.64 & 1,2 & yab,ya & 3 & $\cdots$ & $\cdots$ \\
5 39 37.39 & -69 46  8.81 && 14.573 & 0.067 & 13.769 & 0.083 & 12.907 & 0.083 && 12.92 & 3.51 &  5.36 & 4.80 & $\cdots$ & $\cdots$ &&  73.11 &  7.49 &  93.53 & 19.08 & $\cdots$ & $\cdots$ & 4.82 & 1 & yab & 4 & $\cdots$ & $\cdots$ \\
5 39 35.89 & -69 46  3.32 && 15.451 & 0.043 & 14.408 & 0.058 & 13.405 & 0.051 &&  3.02 & 2.21 & $\cdots$ & $\cdots$ & $\cdots$ & $\cdots$ &&  60.37 & 16.86 &  $\cdots$ & $\cdots$ & $\cdots$ & $\cdots$ & 5.78 & 1 & ya & $\cdots$ & $\cdots$ & $\cdots$ \\
5 39 43.56 & -69 45 39.68 && 18.559 & 0.056 & 17.214 & 0.042 & 15.942 & 0.031 && $\cdots$ & $\cdots$ & 10.70 & 2.86 &  5.70 & 3.15 && $\cdots$ & $\cdots$ & 126.55 &  7.38 & 107.14 & 13.82 & 7.45 & 1,2 & yc,ya & 5 & $\cdots$ & $\cdots$ \\
5 39 59.23 & -69 45 26.02 && 16.326 & 0.025 & 13.882 & 0.013 & 11.771 & 0.004 &&  7.18 & 1.51 &  4.56 & 0.68 &  3.18 & 0.29 && 123.14 &  5.87 & 125.68 &  4.20 & 133.07 & 2.59 & 12.53 & 1,3 & yb,CC5 & $\cdots$ & 1.08 & 3.20 \\
5 39 52.56 & -69 45 16.84 && 13.883 & 0.062 & 13.803 & 0.020 & 13.738 & 0.015 &&  $\cdots$ & $\cdots$ &  2.32 & 1.26 &  1.72 & 1.01 &&  $\cdots$ & $\cdots$ &  98.28 & 13.62 & 109.01 & 14.41 & -0.24 & 1 & yc & $\cdots$ & $\cdots$ & $\cdots$ \\
5 40  9.33 & -69 44 53.59 && 17.918 & 0.045 & 16.830 & 0.052 & 15.931 & 0.067 && 11.67 & 3.60 &  7.60 & 2.87 & $\cdots$ & $\cdots$ &&  13.00 &  8.42 &  77.47 & 10.11 & $\cdots$ & $\cdots$ & 5.19 & 1 & yab & 6 & $\cdots$ & $\cdots$ \\
5 39 45.03 & -69 44 50.32 && 14.693 & 0.039 & 14.309 & 0.043 & 13.942 & 0.054 &&  3.32 & 2.22 & $\cdots$ & $\cdots$ & $\cdots$ & $\cdots$ && 128.82 & 15.88 &  $\cdots$ & $\cdots$ & $\cdots$ & $\cdots$ & 1.60 & 1 & yc & $\cdots$ & $\cdots$ & $\cdots$ \\
5 39 44.35 & -69 44 34.88 && 16.603 & 0.026 & 15.770 & 0.020 & 15.218 & 0.036 &&  $\cdots$ & $\cdots$ &  1.03 & 1.02 & $\cdots$ & $\cdots$ && $\cdots$ & $\cdots$ & 133.88 & 20.09 & $\cdots$ & $\cdots$ & 2.84 & 1,2 & yab,ya & $\cdots$ & $\cdots$ & $\cdots$ \\
5 40  3.46 & -69 43 55.38 && 17.993 & 0.058 & 16.730 & 0.078 & 15.832 & 0.096 && $\cdots$ & $\cdots$ &  5.64 & 4.12 & $\cdots$ & $\cdots$ && $\cdots$ & $\cdots$ &  50.66 & 16.85 & $\cdots$ & $\cdots$ & 5.60 & 1 & ysc & $\cdots$ & $\cdots$ & $\cdots$ \\
5 39 29.75 & -69 43 33.11 && 18.660 & 0.052 & 17.019 & 0.028 & 15.871 & 0.027 && $\cdots$ & $\cdots$ &  7.93 & 2.83 &  9.00 & 3.11 && $\cdots$ & $\cdots$ &  26.76 &  9.60 & 24.00 &  9.34 & 6.64 & 1 & yb & $\cdots$ & $\cdots$ & $\cdots$ \\
5 39 58.23 & -69 47 18.80 && 16.216 & 0.029 & 15.394 & 0.042 & 15.246 & 0.041 && $\cdots$ & $\cdots$ & $\cdots$ & $\cdots$ &  5.92 & 2.70 &&  $\cdots$ & $\cdots$ & $\cdots$ & $\cdots$ &  77.83 & 11.86 & 0.43 & 2 & ya & $\cdots$ & $\cdots$ & $\cdots$ \\
5 39 30.82 & -69 46 48.06 && 17.169 & 0.021 & 16.444 & 0.016 & 16.204 & 0.030 &&  4.32 & 1.55 &  5.38 & 1.21 & $\cdots$ & $\cdots$ && 149.11 &  9.67 & 143.25 &  6.28 & $\cdots$ & $\cdots$ & 0.87 & 2 & ya & $\cdots$ & $\cdots$ & $\cdots$ \\
5 39  3.77 & -69 45 22.15 && 16.770 & 0.011 & 15.802 & 0.015 & 15.508 & 0.022 && $\cdots$ & $\cdots$ &  0.92 & 0.84 &  5.89 & 2.38 && $\cdots$ & $\cdots$ &  95.04 & 19.32 & 158.45 & 10.73 & 1.21 & 2 & ya & $\cdots$ & $\cdots$ & $\cdots$ \\
5 39 40.69 & -69 45 15.91 && 19.334 & 0.090 & 18.145 & 0.040 & 17.358 & 0.074 && $\cdots$ & $\cdots$ &  4.06 & 3.58 & $\cdots$ & $\cdots$ && $\cdots$ & $\cdots$ & 117.82 & 18.93 & $\cdots$ & $\cdots$ & 4.50 & 2 & ya & $\cdots$ & $\cdots$ & $\cdots$ \\
5 39 40.18 & -69 44 54.75 && 16.499 & 0.021 & 16.179 & 0.037 & 16.132 & 0.051 &&  3.15 & 1.30 & $\cdots$ & $\cdots$ &  6.83 & 3.42 &&  89.36 & 10.93 & $\cdots$ & $\cdots$ & 109.46 & 12.8 & -0.36 & 2 & yb & $\cdots$ & $\cdots$ & $\cdots$ \\
5 39 53.15 & -69 44 16.84 && 17.779 & 0.032 & 17.029 & 0.029 & 16.746 & 0.050 &&  7.03 & 2.59 &  3.59 & 2.59 & $\cdots$ & $\cdots$ &&  90.63 &  9.88 & 147.58 & 16.75 & $\cdots$ & $\cdots$ & 1.20 & 2 & yb & $\cdots$ & $\cdots$ & $\cdots$ \\
5 40  0.98 & -69 44  7.32 && 16.495 & 0.015 & 15.962 & 0.028 & 15.926 & 0.043 &&  $\cdots$ & $\cdots$ & $\cdots$ & $\cdots$ &  4.27 & 3.03 && $\cdots$ & $\cdots$ & $\cdots$ & $\cdots$ & 161.77 & 16.54 & -0.38 & 2 & yb & $\cdots$ & $\cdots$ & $\cdots$ \\
5 39 21.16 & -69 44  8.25 && 16.431 & 0.015 & 15.526 & 0.018 & 14.949 & 0.015 &&  1.59 & 0.95 &  1.08 & 0.97 &  1.87 & 1.65 && 147.03 & 14.65 & 138.25 & 19.06 &  87.87 & 18.93 & 2.97 & 2 & ya & $\cdots$ & $\cdots$ & $\cdots$ \\
5 39  3.85 & -69 44  8.54 && 16.599 & 0.008 & 15.986 & 0.011 & 15.890 & 0.021 &&  $\cdots$ & $\cdots$ &  1.59 & 0.84 &  9.24 & 2.61 &&  $\cdots$ & $\cdots$ & 129.36 & 13.35 & 170.73 &  7.76 & -0.04 & 2 & ya & $\cdots$ & $\cdots$ & $\cdots$ \\
5 39 11.01 & -69 43 53.08 && 17.215 & 0.015 & 16.607 & 0.015 & 16.568 & 0.040 &&  2.86 & 1.32 &  1.66 & 1.28 & 14.99 & 4.33 &&  51.84 & 11.98 & 173.32 & 17.46 & 105.16 &  7.93 & -0.37 & 2 & ya & 7 & $\cdots$ & $\cdots$ \\
5 40 15.40 & -69 43  1.28 && 17.103 & 0.013 & 16.313 & 0.015 & 16.082 & 0.033 &&  $\cdots$ & $\cdots$ &  3.72 & 1.08 &  2.85 & 2.69 &&  $\cdots$ & $\cdots$ &   4.49 &  7.99 &  41.81 & 19.63 & 0.83 & 2 & ya & $\cdots$ & $\cdots$ & $\cdots$ \\
5 40 19.37 & -69 43  0.12 && 17.639 & 0.034 & 17.000 & 0.028 & 16.919 & 0.057 && 10.18 & 2.86 & $\cdots$ & $\cdots$ &  9.44 & 5.74 && 107.91 &  7.75 & $\cdots$ & $\cdots$ & 37.36 & 14.86 & -0.06 & 2 & ybc & 8 & $\cdots$ & $\cdots$ \\
5 39 15.01 & -69 42 41.27 && 18.181 & 0.036 & 17.479 & 0.032 & 17.198 & 0.061 &&  9.54 & 3.81 &  4.59 & 2.82 & $\cdots$ & $\cdots$ &&  51.32 & 10.60 & 124.97 & 14.96 & $\cdots$ & $\cdots$ & 1.19 & 2 & yab & $\cdots$ & $\cdots$ & $\cdots$ \\
5 40 15.32 & -69 42 23.43 && 17.192 & 0.036 & 16.383 & 0.053 & 16.187 & 0.056 &&  $\cdots$ & $\cdots$ & $\cdots$ & $\cdots$ & $\cdots$ & $\cdots$ &&  $\cdots$ & $\cdots$ & $\cdots$ & $\cdots$ & $\cdots$ & $\cdots$ & 0.80 & 2 & ya & $\cdots$ & $\cdots$ & $\cdots$ \\
5 39 58.75 & -69 41 10.03 && 17.894 & 0.030 & 17.234 & 0.026 & 17.053 & 0.048 &&  6.85 & 3.21 &  3.07 & 2.19 & $\cdots$ & $\cdots$ &&  56.29 & 12.14 &  28.76 & 16.63 & $\cdots$ & $\cdots$ & 0.54 & 2 & ya & $\cdots$ & $\cdots$ & $\cdots$ \\
5 39 13.78 & -69 40 27.42 && 18.758 & 0.054 & 17.823 & 0.029 & 17.551 & 0.082 && $\cdots$ & $\cdots$ &  9.96 & 3.69 & $\cdots$ & $\cdots$ && $\cdots$ & $\cdots$ & 117.57 &  9.94 & $\cdots$ & $\cdots$ & 1.27 & 2 & ya & $\cdots$ & $\cdots$ & $\cdots$ \\
5 40 26.68 & -69 39 36.28 && 17.108 & 0.023 & 16.472 & 0.012 & 16.313 & 0.053 &&  2.89 & 2.64 &  5.98 & 1.64 & $\cdots$ & $\cdots$ &&  70.31 & 19.30 &  15.67 &  7.55 & $\cdots$ & $\cdots$ & 0.39 & 2 & ya & $\cdots$ & $\cdots$ & $\cdots$ \\
5 39 38.67 & -69 39  3.53 && 17.646 & 0.057 & 15.902 & 0.032 & 14.392 & 0.034 && 27.67 & 6.71 &  8.61 & 2.29 &  8.98 & 2.62 &&  99.32 &  6.74 &  45.25 &  7.34 &  36.49 &  8.01 & 8.94 & 2 & ya & 9 & $\cdots$ & $\cdots$ \\
5 40 47.19 & -69 37  5.67 && 17.111 & 0.022 & 16.248 & 0.008 & 16.166 & 0.059 &&  3.05 & 1.76 &  1.96 & 1.44 & $\cdots$ & $\cdots$ &&  32.53 & 14.31 &  97.95 & 16.98 & $\cdots$ & $\cdots$ & -0.05 & 2 & ya & $\cdots$ & $\cdots$ & $\cdots$ \\
5 39  8.48 & -69 44 14.26 && 16.004 & 0.011 & 15.288 & 0.006 & 15.120 & 0.013 &&  1.58 & 0.87 &  1.26 & 0.43 &  3.42 & 1.40 &&  32.11 & 13.81 & 176.29 &  9.19 &   7.03 & 10.88 & 0.41 & 3 & CC5 & $\cdots$ & 0.78 & 4.52 \\
5 39 37.58 & -69 45 40.16 && 14.308 & 0.007 & 12.975 & 0.007 & 12.346 & 0.003 &&  2.11 & 0.33 &  1.41 & 0.34 &  0.41 & 0.23 && 179.34 &  4.40 & 163.88 &  6.81 & 176.72 & 14.29 & 3.26 & 3 & CC5 & $\cdots$ & 0.58 & 5.04 \\
5 40  1.07 & -69 43 23.96 && 16.759 & 0.013 & 16.111 & 0.015 & 15.990 & 0.026 &&  2.78 & 1.11 &  4.29 & 1.06 & $\cdots$ & $\cdots$ &&  52.67 & 10.56 &  87.22 &  6.85 & $\cdots$ & $\cdots$ & 0.13 & 3 & CC5 & $\cdots$ & 1.46 & 4.87 \\
5 40  2.78 & -69 41 18.75 && 17.397 & 0.012 & 16.659 & 0.016 & 16.570 & 0.036 &&  6.13 & 1.61 &  4.72 & 1.21 & 10.26 & 4.41 &&   2.54 &  7.26 &  96.26 &  7.13 & 126.54 & 11.28 & -0.06 & 3 & CC5 & $\cdots$ & 1.78 & 4.05 \\
5 39 31.19 & -69 36 37.14 && 15.983 & 0.012 & 14.638 & 0.007 & 14.152 & 0.019 &&  4.07 & 1.09 &  1.56 & 0.67 &  3.33 & 1.45 && 129.88 &  7.40 & 110.7 & 11.31 &  14.27 & 11.39 & 2.38 & 3 & CC5 & $\cdots$ & 1.02 & 3.39 \\
5 39 49.48 & -69 38  3.27 && 15.352 & 0.009 & 15.199 & 0.005 & 15.160 & 0.024 &&  1.80 & 0.63 &  2.06 & 0.55 & $\cdots$ & $\cdots$ && 155.72 &  9.49 &  39.01 &  7.35 & $\cdots$ & $\cdots$ & -0.39 & 3 & CC5 & $\cdots$ & 1.70 & 4.45 \\
5 39 52.72 & -69 36 34.69 && 16.876 & 0.015 & 15.784 & 0.003 & 15.491 & 0.028 &&  9.76 & 1.65 &  3.98 & 0.79 &  8.78 & 3.55 && 166.06 &  4.76 &  59.33 &  5.53 & 135.37 & 10.73 & 1.19 & 3 & CC5 & $\cdots$ & 0.78 & 4.01 \\
5 39 59.46 & -69 37 29.89 && 16.900 & 0.027 & 15.178 & 0.021 & 13.274 & 0.011 &&  4.66 & 2.35 &  3.16 & 1.43 & $\cdots$ & $\cdots$ && 173.23 & 12.90 &  79.40 & 11.79 & $\cdots$ & $\cdots$ & 11.28 & 3 & CC5 & $\cdots$ & 1.69 & 3.81 \\
5 40 11.57 & -69 33 15.60 && 13.936 & 0.023 & 13.354 & 0.015 & 13.095 & 0.015 &&  $\cdots$ & $\cdots$ & $\cdots$ & $\cdots$ &  2.16 & 1.25 && $\cdots$ & $\cdots$ & $\cdots$ & $\cdots$ & 105.00 & 14.30 & 0.98 & 3 & CC5 & $\cdots$ & 0.77 & 3.38 \\
5 39 43.23 & -69 38 54.01 && 13.722 & 0.082 & 13.011 & 0.083 & 12.374 & 0.068 && 25.49 & 4.81 & 15.01 & 6.14 &  9.35 & 5.38 &&  11.18 &  5.30 &  29.57 & 10.84 &   8.27 & 14.27 & 3.42 & 4 & HEB & 10 & $\cdots$ & $\cdots$ \\
5 39 45.90 & -69 38 39.32 && 13.640 & 0.087 & 13.017 & 0.068 & 11.156 & 0.078 && $\cdots$ & $\cdots$ & $\cdots$ & $\cdots$ & 52.79 & 6.99 && $\cdots$ & $\cdots$ & $\cdots$ & $\cdots$ & 173.87 &  3.76 & 10.51 & 4 & HEB & 11 & $\cdots$ & $\cdots$ \\
5 39 44.12 & -69 38 33.47 && 15.213 & 0.061 & 14.886 & 0.045 & 14.199 & 0.047 &&  8.81 & 4.93 & 24.38 & 4.82 &  5.18 & 2.85 && 142.24 & 13.98 &  43.44 &  5.55 &  16.10 & 13.80 & 3.52 & 4 & YSO & 12 & $\cdots$ & $\cdots$ \\
\enddata
\tablecomments{Information for each column of the table is described below:\\
Columns (1)$-$(2) Equatorial coordinates (J2000.0) in decimal degrees;\\
Columns (3)$-$(8) $J$, $H$, and $K_{s}$ magnitude and error;\\
Columns (9)$-$(14) $J$, $H$, and $K_{s}$ polarization degree and error in units of percentage;\\
Columns (15)$-$(20) $J$, $H$, and $K_{s}$ polarization position angle and error in units of degrees;\\
Columns (21) Visual extinction in unit of magnitudes;\\
Columns (22) References;\\
Columns (23) Types of the matched objects by the literature;\\
Columns (24) Numbering of highly polarized sources that matched with YSO candidates;\\
Columns (25)$-$(26) Mid-infrared colors from $AKARI$ IRC point-source catalog of the LMC in unit of magnitudes.}
\tablenotetext{a}{Evolutionary stages for YSO candidates are labeled as `ya', `yb', and `yc', indicating their stages from \textrm{I} to \textrm{III}, which refer to \citet{che10} and \citet{car12}. The label `ysc', `HEB', and `YSO' indicate young stellar cluster \citep{che10}, HEB objects and massive YSO from \citet{mar08}, respectively. The label `CC5' indicates YSO candidates with water ice absorption, classified by \citet{kat12}.}
\tablenotetext{b}{For the sources that matched with the YSOs candidates from the literature are listed.}
\tablenotetext{c}{Mid-infrared colors that calculated from $AKARI$ IRC point-source catalog of the LMC \citep{kat12}.}
\tablerefs{(1) \citet{che10}, (2) \citet{car12}, (3) \citet{kat12}, (4) \citet{mar08}.}
\end{deluxetable}
\clearpage

\begin{deluxetable}{cccccccc}
\tabletypesize{\scriptsize}
\tablecaption{Criteria for the best quality polarization in the N159/N160 field \label{tab-2}}
\tablewidth{0pt}
\setlength{\tabcolsep}{0.05in}
\tablehead{\colhead{} & \multicolumn{3}{c}{N159} & \colhead{} & \multicolumn{3}{c}{N160}\\
\cline{2-4} \cline{6-8} \\
\colhead{} & \multicolumn{1}{c}{$J$} & \multicolumn{1}{c}{$H$} & \multicolumn{1}{c}{$K_s$} & \colhead{} & \multicolumn{1}{c}{$J$} & \multicolumn{1}{c}{$H$} & \multicolumn{1}{c}{$K_s$}
}
\startdata
$P/\sigma_P$ & & $>$3 & & & & $>$3 & \\
$P$ & & $<$10 & & & & $<$10 & \\
Mag & $\leq$17.0 & $\leq$16.7 & $\leq$14.5 & & $\leq$16.5 & $\leq$16.2 & $\leq$14.0 \\
Total number & 159 & 157 & 56 & & 93 & 120 & 33 \\
\enddata
\end{deluxetable}

\begin{deluxetable}{ccccccccccccccccccccccccccc}
\tabletypesize{\tiny}
\rotate
\setlength{\tabcolsep}{0.02in}
\tablewidth{0pt}
\tablecaption{Photometric and polarimetric catalog of the sources in the N159/N160 field of the LMC \tablenotemark{a} \label{tab-3}}

\tablehead{\colhead{} & \multicolumn{2}{c}{Position} & \colhead{} & \multicolumn{6}{c}{Magnitude} &\colhead{} & \multicolumn{13}{c}{Polarization Properties} & \colhead{} & \multicolumn{2}{c}{NICER} \\
\cline{2-3} \cline{5-10} \cline{12-24} \cline{26-27}\\

\colhead{Field} & \multicolumn{1}{c}{${{\alpha_{\circ\rm J2000.0}}}$} & \multicolumn{1}{c}{${\delta_{\circ\rm J2000.0}}$} & \colhead{} &
\multicolumn{2}{c}{$J$} & \multicolumn{2}{c}{$H$} & \multicolumn{2}{c}{$K_{s}$} & \colhead{} &
\multicolumn{2}{c}{$P_{J}$} & \multicolumn{2}{c}{$P_{H}$} & \multicolumn{2}{c}{$P_{K_{s}}$} & \colhead{} &
\multicolumn{2}{c}{$\theta_{J}$} & \multicolumn{2}{c}{$\theta_{H}$} & \multicolumn{2}{c}{$\theta_{K_{s}}$} & \colhead{} &
\multicolumn{2}{c}{Av} \\

\colhead{} & \multicolumn{1}{c}{} & \multicolumn{1}{c}{} & \colhead{} &
\multicolumn{2}{c}{(mag)} & \multicolumn{2}{c}{(mag)} & \multicolumn{2}{c}{(mag)} & \colhead{} &
\multicolumn{2}{c}{(\%)} & \multicolumn{2}{c}{(\%)} & \multicolumn{2}{c}{(\%)} & \colhead{} &
\multicolumn{2}{c}{($\arcdeg$)} & \multicolumn{2}{c}{($\arcdeg$)} & \multicolumn{2}{c}{($\arcdeg$)} & \colhead{} &
\multicolumn{2}{c}{(mag)} \\

\colhead{(1)}&\colhead{(2)}&\colhead{(3)}&\colhead{}&
\colhead{(4)}&\colhead{(5)}&\colhead{(6)}&\colhead{(7)}&\colhead{(8)}&\colhead{(9)}&\colhead{}&
\colhead{(10)}&\colhead{(11)}&\colhead{(12)}&\colhead{(13)}&\colhead{(14)}&\colhead{(15)}&\colhead{}&
\colhead{(16)}&\colhead{(17)}&\colhead{(18)}&\colhead{(19)}&\colhead{(20)}&\colhead{(21)}&\colhead{}&
\colhead{(22)}&\colhead{(23)}
}
\startdata
N159 & 5 39 39.31 & -69 47 21.83 && 15.363 & 0.006 & 14.195 & 0.010 & 13.828 & 0.004 && 3.97 & 0.38 & 2.01 & 0.61 & 2.94 & 0.57 && 47.44 & 2.71 & 61.31 & 8.28 & 50.31 & 5.45 && 1.63 & 0.90 \\
N159 & 5 40 13.13 & -69 47 16.40 && 17.320 & 0.013 & 16.545 & 0.022 & 16.305 & 0.100 && $\cdots$ & $\cdots$ & 4.92 & 1.49 & $\cdots$ & $\cdots$ && $\cdots$ & $\cdots$ & 83.32 & 8.30 & $\cdots$ & $\cdots$ && 1.05 & 1.09 \\
N159 & 5 39 43.87 & -69 47 13.21 && 16.756 & 0.013 & 15.539 & 0.013 & 15.046 & 0.016 && 8.02 & 1.16 & 6.00 & 0.80 & 2.54 & 1.55 && 55.28 & 4.11 & 52.22 & 3.80 & 70.53 & 14.92 && 2.44 & 0.90 \\
N159 & 5 39 27.05 & -69 47 11.64 && 16.133 & 0.009 & 14.809 & 0.013 & 14.302 & 0.009 && 4.03 & 0.66 & 3.77 & 0.71 & 2.52 & 0.68 && 28.63 & 4.62 & 48.45 & 5.33 & 31.62 & 7.44 && 2.52 & 0.90 \\
N159 & 5 39 45.63 & -69 47 08.93 && 17.051 & 0.009 & 16.138 & 0.012 & 15.851 & 0.023 && 5.89 & 1.18 & 2.77 & 0.88 & 3.63 & 2.43 && 45.11 & 5.64 & 58.10 & 8.64 & 88.84 & 15.94 && 1.16 & 0.91 \\
N159 & 5 39 24.00 & -69 47 01.85 && 16.601 & 0.011 & 15.954 & 0.014 & 15.759 & 0.022 && 4.82 & 0.90 & 1.73 & 0.91 & 7.64 & 2.95 && 54.51 & 5.27 & 132.06 & 13.39 & 98.95 & 10.3 && 0.59 & 0.91 \\
N159 & 5 40 01.63 & -69 46 55.35 && 13.383 & 0.004 & 12.517 & 0.008 & 12.289 & 0.002 && 1.51 & 0.26 & $\cdots$ & $\cdots$ & 0.27 & 0.27 && 173.08 & 4.79 & $\cdots$ & $\cdots$ & 18.53 & 20.07 && 0.77 & 0.90 \\
N159 & 5 39 25.47 & -69 46 59.48 && 16.882 & 0.015 & 16.060 & 0.009 & 15.807 & 0.021 && 3.60 & 0.92 & 2.20 & 0.64 & 3.83 & 2.16 && 64.98 & 7.11 & 42.07 & 8.02 & 4.73 & 14.04 && 0.94 & 0.91 \\
N159 & 5 39 30.82 & -69 46 48.06 && 17.169 & 0.021 & 16.444 & 0.016 & 16.204 & 0.030 && 4.32 & 1.55 & 5.38 & 1.21 & $\cdots$ & $\cdots$ && 149.11 & 9.67 & 143.25 & 6.28 & $\cdots$ & $\cdots$ && 0.87 & 0.92 \\
N159 & 5 39 50.21 & -69 46 43.60 && 16.487 & 0.013 & 15.801 & 0.010 & 15.611 & 0.020 && 3.75 & 0.90 & $\cdots$ & $\cdots$ & 6.39 & 2.17 && 24.23 & 6.66 & $\cdots$ & $\cdots$ & 103.61 & 9.22 && 0.55 & 0.91 \\
\enddata
\tablecomments{Information for each column of the table is given below:\\
Columns (1) Target field names;\\
Columns (2)$-$(3) Equatorial coordinates (J2000.0) in decimal degrees;\\
Columns (4)$-$(9) $J$, $H$, and $K_{s}$ magnitude and error;\\
Columns (10)$-$(15) $J$, $H$, and $K_{s}$ polarization degree and error in units of percentage;\\
Columns (16)$-$(21) $J$, $H$, and $K_{s}$ polarization position angle and error in units of degrees;\\
Columns (22)$-$(23) Visual extinction and uncertainty in unit of magnitudes.}
\tablenotetext{a}{Only a portion of the catalog is listed in Table~\ref{tab-3}.}
\end{deluxetable}
\clearpage

\clearpage

\clearpage

\clearpage

\clearpage

\clearpage

\clearpage

\clearpage

\clearpage

\clearpage


\begin{thebibliography}{}
\bibitem[Bernard et al.(2016)]{ber16} Bernard, A., Neichel, B., Samal, M. R., et al. 2016, \aap, 592, A77
\bibitem[Bica et al.(1996)]{bic96} Bica, E. L. D., Claria, J. J., Dottori, H., Santos, J. F. C., Jr., \& Piatti, A. E. 1996, \apjs, 102, 57
\bibitem[Bolatto et al.(2000)]{bol00} Bolatto, A. D., Jackson, J. M., Israel, F. P., Zhang, X., \& Kim, S. 2000, \apj, 545, 234
\bibitem[Carlson et al.(2012)]{car12} Carlson, L. R., Sewilo, M., Meixner, M., Romita K. A., \& Lawton, B. 2012, \aap, 542, A66
\bibitem[Chen et al.(2010)]{che10} Chen, C. R., Indebetouw, R., Chu, Y.-h., et al. 2010, \apj, 721, 1206
\bibitem[Chu et al.(1997)]{chu97} Chu Y.-H., Kennicutt, R. C., Snowden, S. L., et al. 1997, \pasp, 109, 554
\bibitem[Cohen et al.(1988)]{coh88} Cohen, R. S., Dame, T. M., Garay, G., Montani, J., Rubio, M. \& Thaddeus, P. 1988, \apj, 331, L95
\bibitem[Cowley et al.(1995)]{cow95} Cowley, A. P., Schmidtke, P. C., Anderson, A. L., \& McGrath, T. K. 1995, \pasp, 107, 145
\bibitem[Crutcher(2012)]{cru12} Crutcher, R. M. 2012, \araa, 50, 29
\bibitem[Dobbs et al.(2014)]{dob14} Dobbs, C. L., Krumholz, M. R., Ballesteros-Paredes, J., et al. 2014, in Protostars and Planets VI, ed. H. Beuther et al. (Tucson, AZ: Univ. Arizona Press), 3
\bibitem[Ezawa et al.(2004)]{eza04} Ezawa, H., Kawabe, R., Kohno, K., Yamamoto, S., \& ASTE Team 2004, \procspie, 5489, 763
\bibitem[Fari\~na et al.(2009)]{far09} Fari\~na, C., Bosch, G., Morrell, N. I., Barb\'a, R. H., \& Walborn, N. R. 2009, \aj, 138, 510
\bibitem[Fukui et al.(2015)]{fuk15} Fukui, Y., Harada, R., Tokuda, K., et al. 2015, \apjl, 807, L4
\bibitem[Fukui et al.(2008)]{fuk08} Fukui, Y., Kawamura, A., Minamidani, T., et al. 2008, \apjs, 178, 56
\bibitem[Fukui et al.(1999)]{fuk99} Fukui, Y., Mizuno, N., Yamaguchi, R., et al. 1999, \pasj, 51, 745
\bibitem[Galametz et al.(2013)]{gal13} Galametz, M., Hony, S., Galliano, F., et al. 2013, \mnras, 431, 1596
\bibitem[Heiles \& Crutcher(2005)]{hei05} Heiles, C. \& Crutcher, R. 2005, in Cosmic Magnetic Fields, ed. R. Wielebinski \& R. Beck (Lecture Notes in Physics, Vol. 664; Berlin: Springer), 137
\bibitem[Henize(1956)]{hen56} Henize, K. G. 1956, \apjs, 2, 315
\bibitem[Henning et al.(1998)]{henn98} Henning, Th., Klein, R., Chan, S. J., et al. 1998, \aap, 338, L51
\bibitem[Heydari-Malayeri et al.(1999)]{hey99} Heydari-Malayeri, M., Rosa, M. R., Charmandaris, V., Daharveng, L., \& Zinnecker, H. 1999, \aap, 352, 665
\bibitem[Heydari-Malayeri \& Testor(1986)]{hey86} Heydari-Malayeri, M. \& Testor, G. 1986, \aap, 162, 180
\bibitem[Hunt \& Whiteoak(1994)]{hunt94} Hunt, M. R., \& Whiteoak, J. B. 1994, \pasa, 11, 68
\bibitem[Ita et al.(2008)]{ita08} Ita, Y., Onaka, T., Kato, D., et al. 2008, \pasj, 60, S435
\bibitem[Johansson et al.(1998)]{joh98} Johansson, L. E. B., Greve, A., Booth, R. S., et al. 1998, \aap, 331, 857
\bibitem[Jones(1989)]{jon89} Jones, T. J. 1989, \apj, 346, 728
\bibitem[Jones et al.(1986)]{jon86} Jones, T. J., Hyland, A. R., Straw, S., et al. 1986, \mnras, 219, 603
\bibitem[Jones et al.(2005)]{jon05} Jones, T. J., Woodward, C. E., Boyer, M. L., Gehrz, R. D., \& Polomski, E. 2005, \apj, 620, 731
\bibitem[Kandori et al.(2006)]{kan06} Kandori, R., Kusakabe, N., Tamura, M., et al. 2006, \procspie, 6269, 159
\bibitem[Kandori et al.(2007)]{kan07} Kandori, R., Tamura, M., Kusakabe, N., et al. 2007, \pasj, 59, 487
\bibitem[Kato et al.(2012)]{kat12} Kato, D., Ita, Y., Onaka, T., et al. 2012, \aj, 144, 179
\bibitem[Kim et al.(1999)]{kim99} Kim, S., Dopita, M. A., Staveley-Smith, L., \& Bessell, M. S. 1999, \aj, 118, 2797
\bibitem[Kim et al.(2016)]{kim16} Kim, J., Jeong, W.-S., Pak, S., Park, W.-K., \& Tamura, M. 2016, \apjs, 222, 2
\bibitem[Kim et al.(2011)]{kim11} Kim, J., Pak, S., Choi, M., et al. 2011, JKAS, 44, 135
\bibitem[Kusakabe et al.(2008)]{ku08} Kusakabe, N., Tamura, M., Kandori, R., et al. 2008, \aj, 136, 621
\bibitem[Kusune et al.(2015)]{kus15} Kusune, T., Sugitani, K., Miao, J., et al. 2015, \apj, 798, 60
\bibitem[Kutner et al.(1997)]{kut97} Kutner, M. L., Rubio, M., Booth, R. S, et al. 1997, \aaps, 122, 255
\bibitem[Kwon et al.(2011)]{kwon11} Kwon, J., Tamura, M., Kandori, R., et al. 2011, \apj, 741, 35
\bibitem[Lada et al.(1994)]{lad94} Lada, C. J., Lada, E. A., Clemens, D. P., \& Bally, J. 1994, \apj, 429, 694
\bibitem[Li et al.(2014)]{li14} Li, H.-B., Goodman, A., Sridharan, T. K., et al. 2014, in Protostars and Planets VI, ed. H. Beuther et al. (Tucson, AZ: Univ. Arizona Press), 101
\bibitem[Lombardi \& Alves(2001)]{lom01} Lombardi, M., \& Alves, J. 2001, \aap, 377, 1203
\bibitem[MacLow \& Klessen(2004)]{mac04} MacLow, M.-M., \& Klessen, R. S. 2004, Rev. Mod. Phys., 76, 125
\bibitem[Mart\'in-Hern\'andez et al.(2008)]{mar08} Mart\'in-Hern\'andez, N. L., Peeters, E., \& Tielens, A. G. G. M. 2008, \aap, 489, 1189
\bibitem[Martin et al.(1992)]{mar92} Martin, P. G., Adamson, A. J., Whittet, D. C. B., et al. 1992, \apj, 392, 691
\bibitem[Martin \& Whittet(1990)]{mar90} Martin, P. G., \& Whittet, D. C. B. 1990, \apj, 357, 113
\bibitem[McKee \& Ostriker(2007)]{mc07} McKee, C. F. and Ostriker, E. C. 2007, \araa, 45, 565
\bibitem[Meixner et al.(2010)]{meix10} Meixner, M., Galliano, F., Hony, S., et al. 2010, \aap, 518, L71
\bibitem[Meixner et al.(2006)]{meix06} Meixner, M., Gordon, K. D., Indebetouw, R., et al. 2006, \aj, 132, 2268
\bibitem[Mizuno et al.(2010)]{miz10} Mizuno, Y., Kawamura, A., Onishi, T., et al. 2010, \pasj, 62, 51
\bibitem[Nagata(1990)]{nagt90} Nagata, T. 1990, \apj, 348, 13
\bibitem[Nagayama et al.(2003)]{nag03} Nagayama, T., Nagashima, C., Nakajima, Y., et al. 2003, \procspie, 4841, 459
\bibitem[Nakajima et al.(2005)]{nak05} Nakajima, Y., Kato, D., Nagata, T., et al. 2005, \aj, 129, 776
\bibitem[Nakajima et al.(2007)]{nak07} Nakajima, Y., Kandori, R., Tamura, M., et al. 2007, \pasj, 59, 519
\bibitem[Ott et al.(2008)]{ott08} Ott, J., Wong, T., Pineda, J. L., et al. 2008, \pasa, 25, 129
\bibitem[Padoan et al.(2004)]{pad04} Padoan, P., Jimenez, R., Juvela, M., \& Nordlund, \r A. 2004, \apjl, 604, L49
\bibitem[Pak et al.(1998)]{pak98} Pak, S., Jaffe, D. T., van Dishoeck, E. F., et al. 1998, \apj, 498, 735
\bibitem[Pineda et al.(2009)]{pin09} Pineda, J. L., Ott, J., Klein, U., Wong, T., Muller, E., \& Hughes, A. 2009, \apj, 703, 736
\bibitem[Points et al.(2001)]{poi01} Points, S. D., Chu, Y.-H., Snowden, S. L., \& Smith, R. C. 2001, \apjs, 136, 99
\bibitem[Preibisch \& Zinnecker(1999)]{pre99} Preibisch, Th. \& Zinnecker, H. 1999, \aj, 117, 2381
\bibitem[Preibisch \& Zinnecker(2007)]{pre07} Preibisch, Th. \& Zinnecker, H. 2007, in IAU Symp. 237, Triggered Star Formation in a Turbulent ISM, ed. B. G. Elmegreen \& J. Palous (Dordrecht: Kluwer)
\bibitem[Saito et al.(2009)]{sai09} Saito, H., Tamura, M., Kandori, R., et al. 2009, \aj, 137, 3149
\bibitem[Santos et al.(2014)]{san14} Santos, F. P., Franco, G. A. P., Roman-Lopes, A., Reis, W., \& Rom\'an-Z\'u\~niga, C. G. 2014, \apj, 783, 1
\bibitem[Seok et al.(2013)]{seok13} Seok, J. Y., Koo, B.-C., \& Onaka, T. 2013, \apj, 779, 134
\bibitem[Serkowski, Mathewson, \& Ford(1975)]{ser75} Serkowski, K., Mathewson, D. S., \& Ford, V. L. 1975, \apj, 196, 261
\bibitem[Seward et al.(2010)]{sew10} Seward, F. D., Williams, R. M., Chu, Y.-H., Gruendl, R. A., \& Dickel, J. R. 2010, \aj, 140, 177
\bibitem[Shimonishi et al.(2008)]{shi08} Shimonishi, T., Onaka, T., Kato, D., et al. 2008, \apj, 686, L99
\bibitem[Shu, Adams, \& Lizano(1987)]{shu87} Shu, F. H., Adams, F. C., \& Lizano, S. 1987, \araa, 25, 23
\bibitem[Skrutskie et al.(2006)]{skr06} Skrutskie, M. F., Cutri, R. M., Stiening, M. D., et al. 2006, \aj, 131, 1163
\bibitem[Stetson(1987)]{ste87} Stetson, P. B. 1987, \pasp, 99, 191
\bibitem[Tamura et al.(2007)]{tam07} Tamura, M., Kandori, R., Hishimoto, J., et al. 2007, \pasj, 59, 467
\bibitem[Testor et al.(2006)]{tes06} Testor, G., Lemaire, J. L., Field, D., \& Diana, S. 2006, \aap, 453, 517
\bibitem[Testor et al.(2007)]{tes07} Testor, G., Lemaire, J. L., Kristensen, L. E., Field, D., \& Diana, S. 2007, \aap, 469, 459
\bibitem[van Loon et al.(2010)]{van10} van Loon, J. T., Oliveira, J. M., Gordon, K. D., et al. 2010, \aj, 139, 68
\end{thebibliography}
\end{document}